\documentclass{article}
\DeclareUnicodeCharacter{2009}{\,}
\DeclareUnicodeCharacter{2005}{\,}
\usepackage{graphicx}           
\usepackage{comment}            
\usepackage{authblk}            
\usepackage{xcolor}             
\usepackage{amsmath}            
\usepackage{makecell}           
\usepackage{threeparttable}     
\usepackage{lscape}             
\usepackage[backend=biber,style=nature,url=false]{biblatex}
\usepackage[margin=1in]{geometry}  

\makeatletter
\renewcommand\paragraph{\@startsection{paragraph}{4}{0pt}%
  {2ex \@plus .5ex \@minus .2ex}%
  {0.5ex \@plus 0ex}%
  {\normalfont\normalsize\bfseries}}
\makeatother

\title{Solid-state laser cooling}
\author[1]{Yang Ding}
\author[1]{Shenghao Zhang}
\author[2]{Alexander R. Albrecht}
\author[3]{Zhaojie Feng}
\author[3]{Lars Forberger}
\author[4]{Hiroki Tanaka}
\author[5]{Markus P. Hehlen}
\author[6]{Galina Nemova}
\author[3]{Peter J. Pauzauskie}
\author[2,6]{Denis V. Seletskiy}
\author[1,7]{Masaru Kuno}
\affil[1]{Department of Chemistry and Biochemistry, University of Notre Dame, Notre Dame, IN 46556, USA}
\affil[2]{Department of Physics and Astronomy, University of New Mexico, Albuquerque, NM 87131, USA}
\affil[3]{Department of Materials Science, University of Washington, Seattle, Washington 98195, USA}
\affil[4]{Leibniz-Institut für Kristallzüchtung (IKZ), Max-Born-Str. 2, 12489 Berlin, Germany}
\affil[5]{Los Alamos National Laboratory, P.O. Box 1663, Los Alamos, NM, 87545, USA}
\affil[6]{Department of Engineering Physics, Polytechnique Montréal, Montréal, QC, H3T 1J4, Canada}
\affil[7]{University of Notre Dame, Department of Physics and Astronomy, Notre Dame, IN 46556, USA}

\begin{document}
\maketitle
\section{Abstract}
Since the first proof-of-concept demonstrations of photoluminescence-based optical refrigeration, solid-state laser cooling has developed into a credible competitor to conventional cryogenic technologies. Solid-state laser cooling continues to advance as new materials push cooling limits. These developments have created a need to consolidate progress made to date as well as standardize critical experimental considerations needed for reliable and verifiable cooling measurements. This primer therefore outlines essential concepts and requirements, which underpin solid-state laser cooling. The primer summarizes key milestones achieved with cooling-grade, rare-earth-doped glasses and crystals as well as with semiconductors. It additionally highlights emerging applications of solid-state optical refrigeration. To strengthen the consistency and reproducibility of cooling results going forward, two reporting checklists are introduced. They cover materials, cooling metrics, and thermometry. This primer is intended to serve as both a tutorial and a practical reference for incoming and existing researchers involved in solid-state laser-cooling.
\section{Introduction}
Nearly sixty years before atomic gases were Doppler cooled, Peter Pringsheim proposed a conceptually simple way to cool matter using photoluminescence (PL).\cite{pringsheim1929zwei} In his gedanken experiment, Pringsheim posited that filtered ($^2$P$_\frac{1}{2}\rightarrow ^2$S$_\frac{1}{2}$, 589.952 nm\cite{risberg1956revision}) emission from a sodium vapor lamp could be used to excite electrons in a neighboring Na vapor from their $^2\textrm{S}
_{\frac{1}{2}}$ ground level to their first $^2$\textrm{P}$_{\frac{1}{2}}$ excited level. Inelastic collisions in the gas would then excite some of the electrons from the first $^2$\textrm{P}$_{\frac{1}{2}}$ excited level to a second excited $^2$\textrm{P}$_{\frac{3}{2}}$ level from where electrons relaxing from $^2$\textrm{P}$_{\frac{3}{2}}$ to $^2$\textrm{S}$_{\frac{1}{2}}$ would emit higher energy/shorter wavelength (588.995 nm\cite{risberg1956revision}), PL. Because their emission possessed an energy greater than that initially absorbed, energy would be removed from the Na vapor. Based on the $^2$P$_{\frac{3}{2}}$-$^2$P$_{\frac{1}{2}}$ level splitting, approximately 3 meV of translational energy would be removed per absorbed/emitted photon. In this way, Pringsheim envisioned how photoluminescence up-conversion could induce refrigeration. Of note is the broader applicability of Pringsheim's concept to cooling condensed phases.

In 1945, Vavilov called into question Pringsheim's idea, arguing that the concept violated a heat engine's Carnot efficiency.\cite{vavilov1945some} The ensuing debate\cite{pringsheim1946some,vavilov1946photoluminescence} was settled by Lev Landau in 1946 when he demonstrated that Pringsheim's PL cooling did not violate the Second Law of thermodynamics.\cite{landau1946thermodynamics} To do this, he considered the entropy of a radiation field based on Bose-Einstein statistics applied to a photon gas. Landau established that radiation's entropy is a function of its angle of propagation and its frequency bandwidth. So long as the increase in radiation entropy exceeded the cooling specimen's entropy decrease, there was no violation of the Second Law.  

Nearly seven decades later, Epstein and co-workers provided experimental proof-of-concept by cooling a rare earth (RE\textsuperscript{3+}) Yb$^{3+}$-doped fluoride glass by 0.3 K below room temperature.\cite{epstein1995observation} Two decades later, Yb$^{3+}$-doped fluoride crystals were cooled to cryogenic temperatures.\cite{seletskiy2010laser,melgaard2016solid} Today, 70 K is within reach and temperatures as low as 50 K are predicted for next generation RE\textsuperscript{3+}-doped crystals, which exhibit 10-fold improvements in purity.\cite{seletskiy2010laser,seletskiy2011spie,volpi2019optical}

RE\textsuperscript{3+} laser cooling has now become a serious competitor to conventional, cryogenic cooling technologies. It offers all-optical, vibration-free cooling in a miniaturaizable format. Case uses include next-generation metrology and sensing applications. RE\textsuperscript{3+} optical refrigeration, however, is not without its limitations. RE\textsuperscript{3+}-doped cooling materials possess global minimum achievable temperatures (gMATs) that result from the thermal depopulation of RE\textsuperscript{3+} ground state multiplets as well as from background absorption-induced heating from impurities. Both ultimately limit how low RE\textsuperscript{3+} materials can be cooled.

In the ongoing push to reach lower optical refrigeration temperatures, new cooling materials are needed. Significant attention has therefore shifted to demonstrating semiconductor optical refrigeration. This has been spurred by calculated semiconductor laser cooling floors as low as 10 K and by their ready integration into electronic and optical devices. 

This primer develops core concepts underlying photoluminescence-based, optical refrigeration. It provides key material and experimental design considerations. It also points to available opportunities where future frontier studies can push current laser cooling limits. In what follows, \textbf{Section \ref{section:experimentation}} discusses materials and key relationships that dictate cooling as well as light sources best suited to realizing solid-state optical refrigeration. \textbf{Section \ref{section:results}} presents recent laser cooling advances in RE\textsuperscript{3+}-doped solids and semiconductors. \textbf{Section \ref{section:applications}} is devoted to applications, focusing on optical cryocoolers, radiation-balanced lasers, and uses in metrology. Because solid-state laser cooling results are sensitive to small variations in material quality, optical coupling, heat management, and thermometry, \textbf{Section \ref{section:reproducibility}} discusses reproducibility and data deposition standards. \textbf{Section \ref{section:limitations}} describes current limitations of solid state laser cooling that remain open challenges for the field. \textbf{Section \ref{section:outlook}} concludes with an outlook.

\section{Experimentation}\label{section:experimentation}
\subsection{Key cooling constraints that dictate material selection and experimental conditions}\label{subsect:cooling_formulas}

\begin{figure}
\centering
\includegraphics[width=0.75\linewidth]{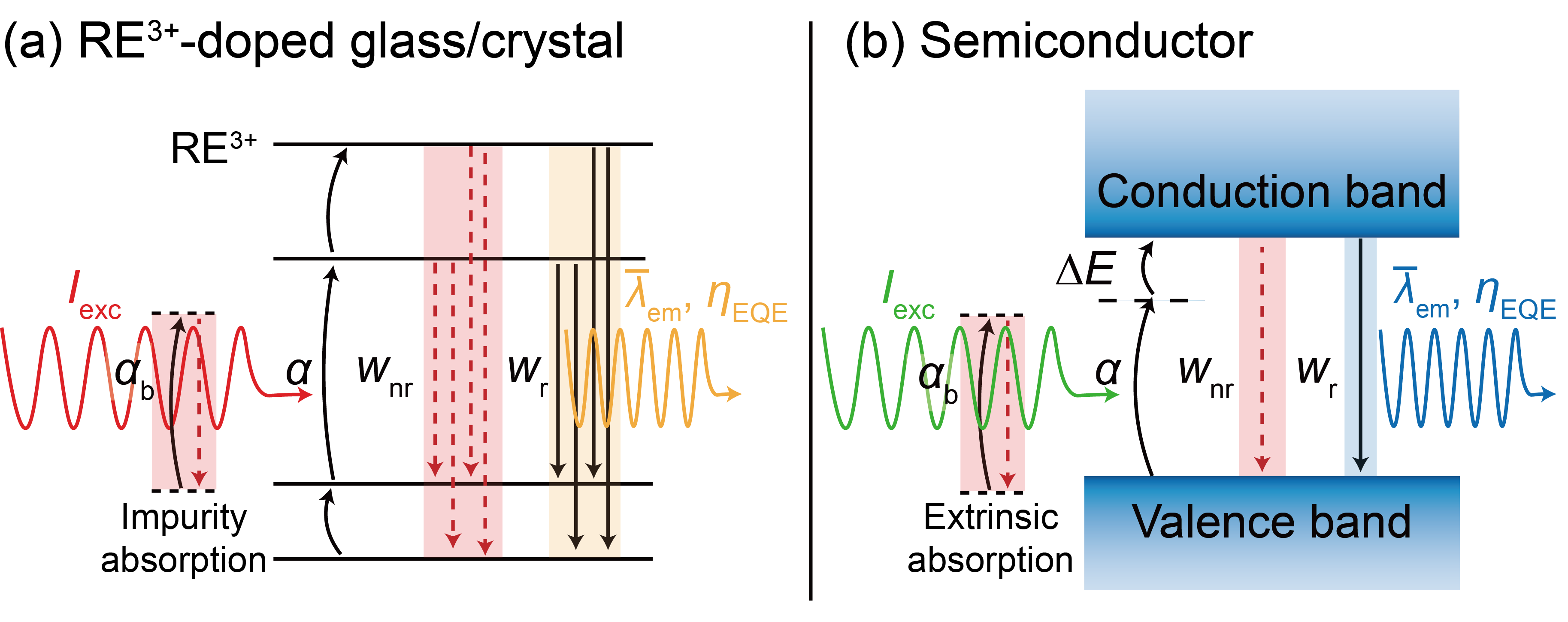}
\caption{Schematic of laser cooling cycles in RE\textsuperscript{3+}-doped glasses/crystals (left) and semiconductors (right).}
\label{fig:Figure_1}
\end{figure}

To undergo optical refrigeration, both material and experimental conditions must ensure a net positive cooling power density, $P_\textrm{cool} $ (W cm$^{-3}$). $P_\textrm{cool}$ arises within the context of competing absorbed and emitted power densities, \textit{i.e.}, $P_\textrm{cool}=P_\textrm{em}-P_\textrm{abs}$ with $P_\textrm{abs}=\left(\alpha+\alpha_\textrm{b}\right)I_\textrm{exc}$ the absorbed, incident laser power density (W cm$^{-3}$, red and green waves, \textbf{Figure \ref{fig:Figure_1}}), $\alpha$ the material’s absorption coefficient (cm$^{-1}$) at wavelength $\lambda_\textrm{exc}$ (nm), $\alpha_\textrm{b}$ an absorption coefficient due to the existence of a background absorption that results in heating, and $I_\textrm{exc}$, an incident laser excitation intensity (W cm$^{-2}$). In the absence of non-linear processes, \textit{e.g.}, for RE$^{3+}$ ion ensembles excited well below saturation, total absorbed powers ($P_\textrm{abs,tot}$, W) can be approximated by multiplying $P_\textrm{abs}$ with the specimen's illuminated volume. $P_\textrm{abs,tot}$ is also obtained as the product of the specimen absorptance, $A(\lambda_{\rm exc})$ and the incident excitation power $P_\textrm{exc}$ (W), \textit{i.e.}, $P_\textrm{abs,tot}=A(\lambda_{\rm exc})P_\textrm{exc}$.

An accompanying, emitted PL power density is $P_\textrm{em}=\left(\eta_\textrm{EQE} \eta_\textrm{abs}\frac{\lambda_\textrm{exc}}{\bar\lambda_\textrm{em}} \right)P_\textrm{abs}$ where $\bar\lambda_\textrm{em}$ is the mean emission wavelength (yellow and blue waves, \textbf{Figure \ref{fig:Figure_1}}), and $\lambda_\textrm{exc}>\bar\lambda_\textrm{em}$ indicates the anti-Stokes photoluminescence (ASPL) regime where laser cooling becomes possible. $\eta_\textrm{EQE}=\frac{\eta_\textrm{e}W_\textrm{r}}{\eta_\textrm{e}W_\textrm{r}+W_\textrm{nr}}$ is the material's external quantum efficiency where $W_\textrm{r}$ and $W_\textrm{nr}$ are radiative (solid black downward arrows, \textbf{Figure \ref{fig:Figure_1}}) and non-radiative (dashed red downward arrows, \textbf{Figure \ref{fig:Figure_1}}) excited state decay rates, $\eta_\textrm{e}$ is a photon extraction efficiency, and $\eta_\textrm{abs}={\alpha\over \alpha+\alpha_\textrm{b}}$ is an absorption efficiency. In some studies\cite{morozov2017defect,morozov2017photoluminescence,zhang2024resonant,abbas2025efficient,Zhang2025Principles}, $\eta_\textrm{abs}$ is referred to as an up-conversion efficiency, which describes the probability that a below-gap absorption event leads to ASPL.

When $P_\textrm{em}$ and $P_\textrm{abs}$ are combined, the following $P_\textrm{cool}$ expression results,
\begin{equation}
P_\textrm{cool}=\left(\eta_\textrm{EQE} \eta_\textrm{abs}\frac{\lambda_\textrm{exc}}{\bar\lambda_\textrm{em}} -1\right)P_\textrm{abs}.
\label{eqn:Pcool_equation_wavelength}
\end{equation}
From this, a cooling efficiency, $\eta_\textrm{c}$, emerges and is
\begin{equation}
\eta_\textrm{c}=\frac{P_\textrm{cool}}{P_\textrm{abs}}=\eta_\textrm{EQE} \eta_\textrm{abs}\frac{\lambda_\textrm{exc}}{\bar\lambda_\textrm{em}} -1.
\label{eqn:eta_cool_equation_wavelength}
\end{equation}
Alternative expressions in terms of energies or frequencies are
\begin{equation}
P_\textrm{cool}=\left(\eta_\textrm{EQE} \eta_\textrm{abs}\frac{\bar{E}_\textrm{em}}{\bar{E}_\textrm{em}-\Delta E} -1\right)P_\textrm{abs}=\left(\eta_\textrm{EQE}\eta_\textrm{abs}{\bar\nu_\textrm{em}\over\nu_\textrm{exc}} -1\right)P_\textrm{abs},
\label{eqn:Pcool_equation_energy}
\end{equation}
and
\begin{equation}
\eta_\textrm{c}=\eta_\textrm{EQE} \eta_\textrm{abs}\frac{\bar{E}_\textrm{em}}{\bar{E}_\textrm{em}-\Delta E}-1=\eta_\textrm{EQE}\eta_\textrm{abs}{\bar\nu_\textrm{em}\over\nu_\textrm{exc}}-1
\label{eqn:eta_cool_equation energy}
\end{equation}
with $\Delta E = \bar{E}_{\textrm{em}}-E_{\textrm{exc}} = h\bar\nu_\textrm{em}-h\nu_\textrm{exc}$ and with $\bar E_\textrm{em}$ ($\bar\nu_\textrm{em}$) the mean emission energy (frequency).

In all cases, cooling exists for $P_\textrm{cool}>0$. Alternatively stated, $\eta_\textrm{c}$ must be positive. Both lead to stringent requirements on material properties best seen when heating and cooling powers are balanced, \textit{i.e.}, $P_\textrm{cool}=0$. In short, cooling requires that $\eta_\textrm{EQE}\eta_\textrm{abs}$ satisfy $\eta_\textrm{EQE}\eta_\textrm{abs}>\frac{\bar\lambda_\textrm{em}}{\lambda_\textrm{exc}}$.

To put this material constraint into focus, for a $\Delta E=25$ meV $\approx202$ cm$^{-1}$ detuning of the laser excitation energy below a mean emission energy of $\bar E_\textrm{em}=1.24$ eV $=10000$ cm$^{-1}$ ($\bar\lambda_\textrm{em}=1000$ nm), $\eta_\textrm{EQE}\eta_\textrm{abs}>0.98$. Alternatively, for a $\Delta E=100$ meV $\approx807$ cm$^{-1}$ detuning below a mean emission energy of $\bar E_\textrm{em}=2.44$ eV $=19679$ cm$^{-1}$ ($\bar\lambda_\textrm{em}=508$ nm), $\eta_\textrm{EQE}\eta_\textrm{abs}>0.96$. In the best-case scenario where $\eta_\textrm{abs}=1$, a critical (minimum) external quantum yield (QY) required to cool is
\begin{equation}
\eta_\textrm{EQE,crit}=1-{\frac{\Delta E}{\bar{E} _{\textrm{em}}}}.
\label{eqn:critical_qy}
\end{equation}
Consequently, $\eta_\textrm{EQE,crit}=0.98$ ($\eta_\textrm{EQE,crit}=0.96$) for the $\Delta E=25$ meV [with $\bar{E} _{\textrm{em}} = 1.24$ eV]($\Delta E=100$ meV, with [$\bar{E} _{\textrm{em}} = 2.44$ eV]) detuning example.

\textbf{Equation \ref{eqn:critical_qy}} simultaneously illustrates how a material's energy gap impacts its propensity to cool. For $\Delta E=100$ meV, $\eta_\textrm{EQE,crit}$ decreases from 0.96 to 0.92 when $\bar{E}_\textrm{em}$ decreases from 2.44 eV to 1.24 eV. This illustrates why materials with smaller $\bar{E}_\textrm{em}$ are preferred for cooling since they relax $\eta_\textrm{EQE,crit}$. Even then, $\eta_\textrm{EQE,crit}$-values for most systems exceed 0.95. \textbf{Table \ref{table:eta_EQE_crit}} summarizes calculated $\eta_\textrm{EQE,crit}$-values at 300 K for various cooling materials at two different laser detunings ($\Delta E=25$ meV and $\Delta E=100$ meV). For both, $\eta_\textrm{abs}=1$ is assumed.
\begin{table}[h]
\begin{center}
\caption{$\eta_\textrm{EQE,crit}$-values calculated from \textbf{Equation \ref{eqn:critical_qy}} for cooling grade materials at 300 K and in the limit that $\eta_\textrm{abs}=1$.}
\begin{tabular}{cccc}
\hline
Material  &$\bar{E}_{\textrm{em}}$ (eV) at 300 K& \makecell{$\eta_\textrm{EQE,crit}$\\$\Delta E=25$\,meV} &
           \makecell{$\eta_\textrm{EQE,crit}$\\$\Delta E=100$\,meV} \\
\hline
\hline
Yb$^{3+}$-doped &1.23$\pm$0.02 \cite{bowman2000new}& 
0.979-0.980 & 0.917-0.920\\
Er$^{3+}$-doped&0.805-2.13 \cite{condon2009optical,fernandez2006anti}&0.969-0.983&0.876-0.931\\
Tm$^{3+}$-doped &0.678-0.692 \cite{hoyt2003laser, patterson2008anti,rostami2019tm,nobel2025optical}& 0.963-0.964 & 0.852-0.855\\
Ho$^{3+}$-doped&0.615 \cite{rostami2019observation}&0.959  &0.837 \\
\makecell{GaAs/GaInP }&1.65-1.91 \cite{Imangholi2005_ELO}& 0.985-0.987 & 0.939-0.948\\
CdSe nanocrystals  &1.97-2.38 \cite{qu2002control}& 0.987-0.989 & 0.949-0.958\\
CsPbBr$_3$ nanocrystals &2.38-2.58 \cite{Brennan2017Origin}& 0.989-0.990 & 0.958-0.961\\
\hline
\end{tabular}
\label{table:eta_EQE_crit}
\end{center}
\end{table}

Realistic modeling of $\eta_\textrm{c}$ must take into account the $\nu$ and $T$ dependencies of $\eta_\textrm{EQE}(T)$, $\eta_\textrm{abs}(\nu,T)$, and $\bar\nu_\textrm{em}(T)$ in \textbf{Equation ~\ref{eqn:eta_cool_equation energy}}. Although $\eta_\textrm{EQE}(T)$ possesses a weak temperature dependence\cite{hoyt2003laser,hehlen2007model,seletskiy2016laser,puschel2021temperature}, which can often be ignored, $\eta_\textrm{abs}$ and $\bar\nu_\textrm{em}$ exhibit strong temperature dependencies seen through exponentially decreasing $\alpha$ with decreasing $T$ due to the thermal depopulation of RE\textsuperscript{3+} ground-state multiplets. $\eta_\textrm{c}$ therefore rapidly approaches zero with decreasing temperature and eventually crosses over into a heating regime when $\eta_\textrm{c}<0$. This crossover temperature is called the minimum achievable temperature (MAT), which is a pump frequency-dependent value, \textit{i.e.}, MAT($\nu$). 

There also exists a unique pump frequency, $\nu_\textrm{opt}$, where a \textit{global} minimum achievable temperature (gMAT) exists across all $\nu$. Formally, gMAT is found by setting $d\eta_\textrm{c}(\nu,T)/d\nu=0$.\cite{seletskiy2011oe} In practice, \textbf{Equation ~\ref{eqn:eta_cool_equation energy}} can be used to calculate gMAT, provided that $\alpha(\nu,T)$ and $\bar\nu_\textrm{em}(\nu,T)$ are measured independently.\cite{seletskiy2013precise} This is how gMAT-values of 190 K for ZBLANP:Yb\textsuperscript{3+} and 115 K for YLF:Yb\textsuperscript{3+} have been estimated.\cite{seletskiy2011oe} In either case, a frequency- and temperature-independent $\alpha_\textrm{b}$ has been assumed. Recent cooling results from Yb\textsuperscript{3+}/Tm\textsuperscript{3+} co-doped crystals moreover show that $\alpha_\textrm{b}$'s temperature dependence must be taken into account, particularly for $T<100$ K.\cite{hehlen2007model,volpi2019optical}

\subsubsection{RE\textsuperscript{3+}-doped materials}\label{subsect:RE3_materials}
\begin{figure}[htb]
\centering
\includegraphics[width=0.9\linewidth]{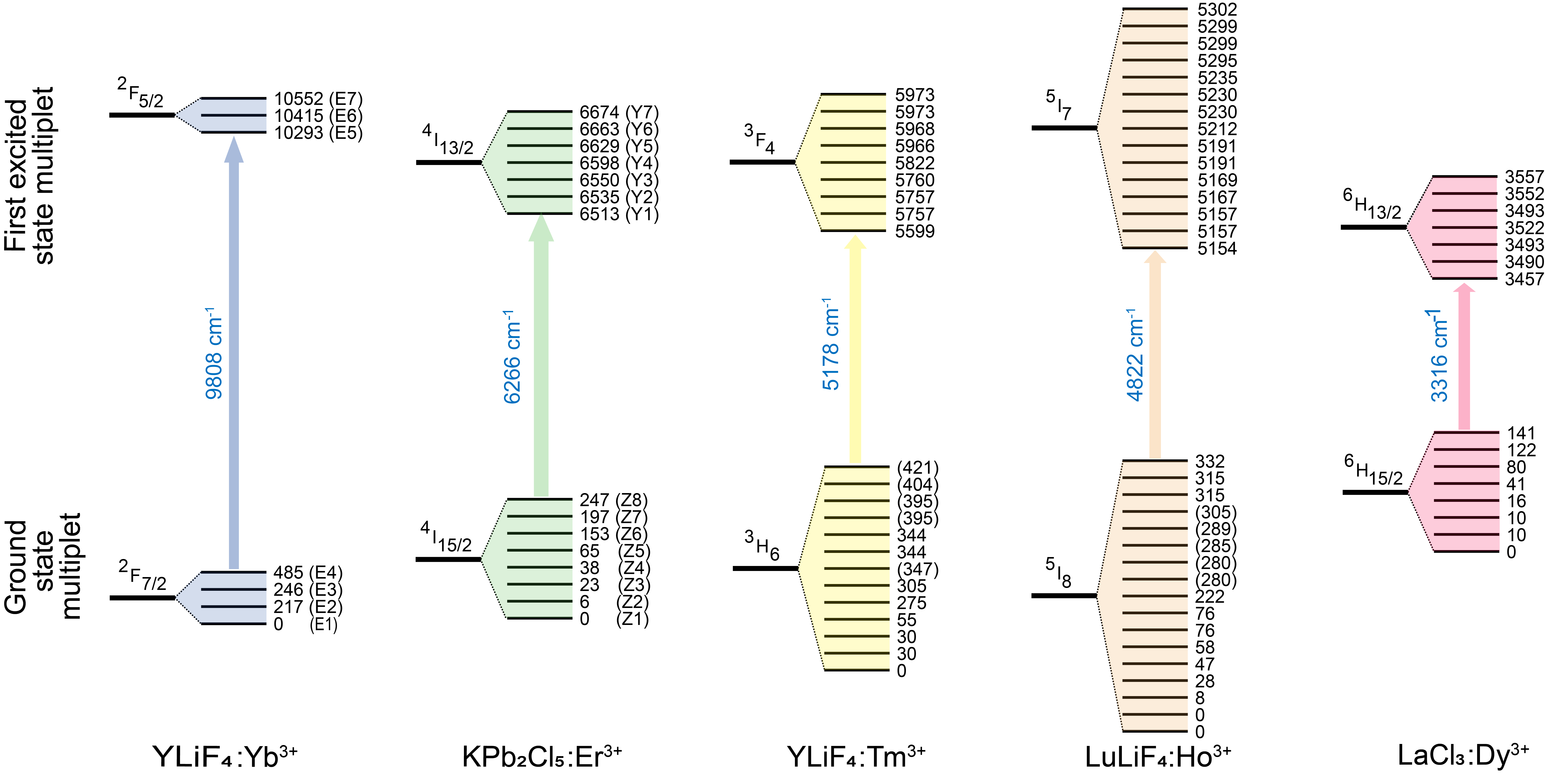}
\caption{Crystal-field energies (in wavenumbers relative to the crystal-field ground state, not to scale) for the ground and first excited state multiplets of Yb\textsuperscript{3+} \cite{demirbas2021detailed,puschel2021temperature}, Er\textsuperscript{3+} \cite{gruber2006modeling}, Tm\textsuperscript{3+} \cite{klimin2010high}, Ho\textsuperscript{3+} \cite{walsh2004spectroscopy}, and Dy\textsuperscript{3+} \cite{rana1988optical} in crystal hosts relevant to solid-state laser cooling. Energies in parenthesis are calculated. The arrows show $E_\textrm{exc}$ for the lowest-energy crystal-field excitations.} 
\label{fig:Figure_2}
\end{figure}

RE\textsuperscript{3+} ion optical properties are governed by transitions between electronic states of their partially filled 4\textit{f} orbitals. This makes them attractive laser-cooling ions as their 4\textit{f} electrons are shielded from the local environment by radially extended and filled 5\textit{s} and 5\textit{p} orbitals.\cite{joos1939,freed1940spectra,kastler1950quelques,hehlen201350th} 4\textit{f}$\longleftrightarrow$4\textit{f} electronic transitions therefore only weakly couple to vibrational modes of the surrounding host matrix. Non-radiative relaxation rates ($W_\textrm{nr}$) are consequently suppressed relative to corresponding radiative rates ($W_\textrm{r}$), in turn, enabling certain RE\textsuperscript{3+} excited states to decay radiatively with high internal quantum efficiencies. \textbf{Table \ref{table:RE3+_summary}} summarizes attractive RE\textsuperscript{3+} ion choices and excited states. \textbf{Figure \ref{fig:Figure_2}} illustrates their ground-state and first excited-state multiplet crystal-field energies in crystals relevant to solid-state laser cooling. 

\begin{table}[h]
\begin{center}
\caption{Cooling RE\textsuperscript{3+} ion excited states and emission energies.}
\begin{tabular}{ccc}
\hline
RE\textsuperscript{3+} & Excited state & $\bar{E}_\textrm{em}$ (eV) (cm$^{-1}$)\\
\hline
\hline
Yb\textsuperscript{3+} & \textsuperscript{2}F\textsubscript{5/2} & 1.23 ($\approx 9920$)\\
Er\textsuperscript{3+} & \textsuperscript{4}I\textsubscript{13/2} & 0.805 ($\approx 6490$)\\
Tm\textsuperscript{3+} & \textsuperscript{3}F\textsubscript{4} & 0.685 ($\approx5525$)\\
Ho\textsuperscript{3+} & \textsuperscript{5}I\textsubscript{7} & 0.615 ($\approx 4960$)\\
Dy\textsuperscript{3+} & \textsuperscript{6}H\textsubscript{13/2} & 0.429 ($\approx 3460$)\\
\hline
\end{tabular}
\label{table:RE3+_summary}
\end{center}
\end{table}

RE\textsuperscript{3+} quantum yields are influenced by their host. This is because multi-phonon relaxation rates are determined by both the RE\textsuperscript{3+} electronic structure and the host material through the energy-gap law, which empirically describes the exponential dependence of $W_\textrm{nr}$ with the minimum number of vibrational quanta, $m={\bar{E}_\textrm{em}\over\hbar\omega_\textrm{max}}$, created during non-radiative relaxation\cite{riseberg1968multiphonon}
\begin{equation}
W_\textrm{nr}=\beta e^{-\gamma m}.
\label{eqn:energy_gap_law}
\end{equation}
In \textbf{Equation \ref{eqn:energy_gap_law}}, $\beta$ and $\gamma$ are host material constants and $\hbar\omega_\textrm{max}$ is the highest-energy vibrational mode responsible for multi-phonon relaxation in the RE\textsuperscript{3+} ion's first coordination shell. As a rule of thumb, radiative relaxation dominates multi-phonon relaxation for $m>5$.\cite{henderson1989, shi2022phonon} This enables applications such as solid-state lasers, phosphors, and luminescence thermometers. $m>8$, however, may be required for applications exceptionally sensitive to multi-phonon relaxation. This includes solid-state laser cooling.\cite{hehlen2009design} 

$W_\textrm{nr}$ is minimized to maximize $\eta_\textrm{EQE}$ by choosing a RE\textsuperscript{3+} ion with an $\bar{E}_\textrm{em}$ (\textbf{Table \ref{table:RE3+_summary}}) that, in combination with a host material's low $\hbar\omega_\textrm{max}$, maximizes $m$. Fluorides ($\hbar\omega_\textrm{max}\approx 350-590$ cm\textsuperscript{-1}) and oxides ($\hbar\omega_\textrm{max}\geq 550$ cm\textsuperscript{-1}) are therefore potentially suitable materials for Yb\textsuperscript{3+}, Er\textsuperscript{3+}, Tm\textsuperscript{3+}, and Ho\textsuperscript{3+} solid-state laser cooling. Hosts with lower $\hbar\omega_\textrm{max}$ (\textit{e.g.}, chlorides, $\hbar\omega_\textrm{max}\approx 200-370$ cm\textsuperscript{-1}) are required for Dy\textsuperscript{3+}.

\textit{Additional host material considerations.} Further host material considerations include the host's chemical inertness and mechanical ruggedness as well as large thermal conductivities and known coefficients of thermal expansion. This will be important for the success of applications outlined in \textbf{Section \ref{section:applications}}. Here, fluorides and oxides are generally preferred over heavy halides (\textit{i.e.}, chlorides, bromides, iodides), which are hygroscopic and prone to degradation. Fluorides have the added advantage of possessing relatively low refractive indices ($n\approx 1.4-1.5$ for $\lambda=2.0-1.0$ $\mu$m), which enhance $\eta_\textrm{e}$.

The next area of critical concern is the host material's purity. Any species that absorbs laser light at $\lambda_\textrm{exc}$ but subsequently decays non-radiatively deposits $\sim E_\textrm{exc}$ of heat into the material. This is usually much greater than the $\sim k_\textrm{B}T$ of energy extracted per cooling cycle. For all practical purposes, the entire emission energy is converted into heat. Consequently, background impurity absorption, as captured by $\alpha_\textrm{b}$, imparts exquisite purity requirements on the laser cooling material. Trace impurities at even the few parts per million (ppm) level severely degrade if not prevent cooling.\cite{hehlen2007model, melgaard2014identification} High-purity precursors and contamination-free growth are therefore critical to producing cooling-grade materials. In this regard, single-crystal growth is a purification process that excludes impurities during a melt's slow unidirectional solidification. By contrast, glasses are made by rapidly quenching a melt, and therefore contain all impurities introduced by their precursors. Whether glass or crystal, surface contamination leads to optically-induced heating and must be minimized via careful sample preparation and handling.

Inhomogeneous broadening is another key consideration when choosing between amorphous and crystalline RE\textsuperscript{3+} host materials.\cite{seletskiy2010laser} Glasses exhibit broad distributions of local coordination geometries about RE\textsuperscript{3+} ions. This leads to distributions of resonant absorption energies. The resulting, temperature-independent inhomogeneous broadening typically exceeds an absorption's temperature-dependent, homogeneous broadening, even at room temperature.\cite{lei1998determination} A pump laser with a linewidth smaller than the inhomogeneous broadening therefore only excites a fraction of available RE\textsuperscript{3+} ions. This limits $P_\textrm{abs}$.

In contrast, a crystal's local coordination geometry is near identical for all RE\textsuperscript{3+} ions. This suppresses inhomogeneous broadening and increases $P_\textrm{abs}$. Narrow resonances stand as one reason why RE\textsuperscript{3+}-doped single crystals such as YLF:Yb\textsuperscript{3+} have cooled to cryogenic temperatures, reaching 87 K.\cite{gragossian2017optical, volpi2019optical} Disordered RE\textsuperscript{3+}-doped glasses, by contrast, have not been cooled as extensively. 

\textit{RE\textsuperscript{3+}-doped crystal/glass morphologies.} The morphology and physical size of laser cooling materials must also be considered. Since the late 2000s, bulk single crystals have dominated the development of high-power, solid-state optical refrigerators, aimed at macroscopic payload cooling and the development of radiation balanced lasers (RBLs) as well as radiation balanced amplifiers (RBAs). These experiments typically use crystallographically-oriented samples and polarized excitation to maximize $P_\textrm{abs}$.

Single-mode optical fibers represent another laser-cooling platform of interest to RBLs and RBAs. These applications use optical cooling to offset heat generated by the quantum defect of the stimulated emission process. Achieving an exact balance between heating and cooling, however, is difficult\cite{knall2021radiation} because of the different spatial evolution of heating and cooling power densities along longitudinal (due to pump light being gradually absorbed) and transverse (due to the non-uniform power density distribution of optical modes) directions of a gain fiber.

Finally, laser-cooling materials may be in the form of $\mu$m- or nm-sized RE\textsuperscript{3+}-doped particles. Research in this area has been motivated by applications of optically levitated nanoparticles for ultra-sensitive force sensors\cite{yin2013optomechanics} wherein optical refrigeration of the nanoparticle counteracts heating induced by a trapping laser.\cite{luntz2021laser} This reduces thermal noise, increases readout sensitivity, and prevents sensor degradation.

Challenges exist when working with nanoscale materials. Simply viewing a nanocrystal as a fragment of the bulk neglects various processes that become active at the nanoscale. To illustrate, nanocrystal QYs are often significantly lower than those of their bulk counterparts.\cite{shi2022phonon} This stems from the introduction of defects due to the abrupt termination of nanocrystal lattices. 

Nanocrystal surfaces are consequently passivated with organic ligands to suppress defect formation. These ligands also help control a nanocrystal's size, morphology, and may also influence RE\textsuperscript{3+} photophysics because of their close proximity to incorporated RE\textsuperscript{3+} dopants. In this regard, energy transfer to ligand vibrational modes may introduce new ligand-mediated, non-radiative relaxation pathways for RE\textsuperscript{3+} ions not present in the bulk. Nanocrystal surface ligands may also produce crystal structures different from the bulk.\cite{uchino1989dependence} Together with structural disorder near nanocrystal surfaces\cite{shi2022phonon}, this can alter host crystal-field splittings, vibrational frequencies, and extent of inhomogeneous broadening. All dictate realizable $P_\textrm{cool}$.

Finally, reducing nanocrystal sizes can engineer phonon densities of state.\cite{shi2022phonon} As illustration, at low temperatures, acoustic phonons play an important role in ensuring energy conservation during multiphonon decay.\cite{meltzer2000electron} Suppressing available acoustic phonons via reductions of the nanocrystal size can therefore introduce a phonon bottleneck that suppresses non-radiative relaxation. This can then promote laser cooling to ultracold temperatures.\cite{nemova2012photnorth,werschler2016coupling} Abovementioned nanoscale complexities thus motivate ongoing research into nanostructure-based laser cooling.

\subsubsection{Semiconductors (Bulk and nanocrystals)}\label{subsect:cooling_materials}
Unlike RE\textsuperscript{3+}-doped systems, semiconductors do not experience thermal depopulation of ground electronic states involved in cooling transitions. This is because their cooling cycles involve valence-to-conduction band transitions and because Fermi statistics guarantee populated valence bands at all temperatures. Cooling floors as low as $\sim 10$ K\cite{PhysRevLett.92.247403} (and potentially lower\cite{Duan2025ZeroPointPhonons}) are therefore possible. Attractive about semiconductors are large $\alpha$-values and ns carrier recombination lifetimes, which stand in contrast to ms rare-earth dopant timescales. Semiconductor $P_\textrm{cool}$-values are therefore intrinsically larger than those of RE\textsuperscript{3+}-doped glasses/crystals.\cite{PhysRevLett.92.247403}

Key features required of cooling grade semiconductors are near-identical to those of RE\textsuperscript{3+}-doped glasses/crystals. Namely, needed are unity/near-unity $\eta_\textrm{EQE}$ and unity/near-unity $\eta_\textrm{abs}$. Both are aided by small $\bar{E}_\textrm{em}$-values as well as by minimal $\alpha_\textrm{b}$-values that adversely impact $\alpha_\textrm{abs}$. Additional semiconductor-specific features that promote cooling are intrinsic absorption/emission Stokes shifts\cite{Brennan2017Origin,Brennan2017Existence,
brennan2020universal} that minimize photon recycling and large electron-phonon coupling constants, $g$, or related Huang-Rhys parameters that yield large $\eta_\textrm{abs}$. \textbf{Table \ref{table:semiconductors}} summarizes literature instances where specific cooling requirements have been met.

\begin{table}[h]
\begin{center}
\caption{Compilation of experimentally measured and known cooling grade semiconductor metrics.}
\begin{tabular}{cccc}
\hline
Material & $\eta_\textrm{EQE}$ & $\eta_\textrm{abs}$ & $g$ (meV) \\
\hline
\hline
GaAs/GaInP double heterostructure& $ 0.995\pm0.001$\cite{Bender2013_GaAsGaInP} & $ 0.901$\cite{Bender2013_GaAsGaInP}& -\\ 
CdSe/CdS nanocrystals & 0.99\cite{Hua2020Optical,Ye2021Phonon}& -&11-12\cite{norris1996size,Ye2021Phonon}\\
CsPbBr$_3$ nanocrystals & $0.96\pm 0.05$\cite{Ding2023Mixed,Roman2020Optically}& 0.99\cite{zhang2024resonant}& 42-100\cite{ramade2018exciton,iaru2017strong,saran2017giant,shinde2017low,cho2022exciton}\\
\hline
\end{tabular}
\label{table:semiconductors}
\end{center}
\end{table}

Of note is CsPbBr\textsubscript{3}'s near-unity $\eta_\textrm{abs}$-value, which exceeds that for GaAs. This points to sizable electron-phonon couplings in perovskite materials. Large $g$-values, in turn, enhance absorption efficiencies (or up-conversion efficiencies). For CsPbBr\textsubscript{3}, what result are unity $\eta_\textrm{abs}$-values within error.\cite{zhang2024resonant} Underscoring the importance of this, although the GaAs specimen in \textbf{Table \ref{table:semiconductors}} possesses an $\eta_\textrm{EQE}$-value above a critical quantum yield required to cool, no cooling has been observed due to its $\eta_\textrm{abs}=0.901$ value being too small to engender $P_\textrm{cool}>0$.\cite{Bender2013_GaAsGaInP} Only transient cooling has been reported on a sub-nanosecond timescale.\cite{Lippmann2019CLEO}

Accompanying near-unity, quantum dot $\eta_\textrm{EQE}$-values are a consequence of careful surface engineering. For II-VI nanocrystals such as CdSe, surface passivation strategies have primarily involved depositing a thin layer of a larger band gap semiconductor atop the nanocrystal.\cite{hines1996synthesis,dabbousi1997cdse,peng1997epitaxial} Examples include CdS or ZnS overcoated CdSe quantum dots (\textit{i.e.}, CdSe/CdS or CdSe/ZnS core/shell nanocrystals). This introduces effective, Type-I band offsets at core/shell interfaces, which confine carriers to the nanocrystal core and suppress any surface-mediated, non-radiative recombination. What result are $\eta_\textrm{EQE}$-values as large as unity.\cite{Ye2021Phonon}

$\eta_\textrm{EQE}$-values of overcoated quantum dots, however, are critically sensitive to the quality of their core/shell interfaces. Because core and shell materials generally exhibit lattice mismatches, epitaxial growth is often not possible. This makes attaining unity/near-unity QYs challenging. The situation is to be contrasted to lead-based, all-inorganic perovskite nanocrystals, where an intrinsic defect tolerance\cite{huang2017lead,kang2017high,kovalenko2017properties} translates to unity/near-unity $\eta_\textrm{EQE}$-values without significant surface modifications.

Another important advantage of colloidal nanocrystals is their size. Because nanocrystal dimensions are typically of order 5-10 nm, they are much smaller than the wavelength of light. Their emission is therefore point-dipole like. This obviates the need for light extraction layers to maximize $\eta_\textrm{e}$ ($\eta_\textrm{EQE}$) unlike in bulk or quantum well semiconductors where large refractive index differences with a surrounding medium lead to low $\eta_\textrm{e}$-values. Of note is that RE\textsuperscript{3+} materials are less prone to this issue due to their smaller refractive indices. Phonon confinement in small nanocrystals may additionally increase the likelihood of phonon-assisted, below gap absorption, in turn, enhancing $\eta_\textrm{abs}$.\cite{werschler2016coupling}

Finally, colloidal nanocrystals possess band edge, absorption/emission Stokes shifts, which reduce reabsorption/photon recycling. For systems such as CdSe or CsPbBr$_3$, Stokes shifts range from 20-100 meV at both room and low temperature.\cite{kuno1997band,Brennan2017Existence, Brennan2017Origin,brennan2020universal} These Stokes shifts are size-dependent and increase with decreasing nanocrystal size. In most nanocrystal systems, observed Stokes shifts are attributed to the existence of ``dark" excitons, which emerge from band edge optical fine  structure.\cite{efros1996band,kuno1997band,efros2000electronic}

\subsection{Light source selection}\label{subsect:light_source_selection}
Principal requirements for the light source in optical refrigeration measurements include power stability at the 0.1–1\% level, low M$^2$-values that allow tight focusing, and polarization control. The latter is essential for structurally anisotropic materials, which exhibit polarization-dependent absorption and emission. Pump linewidths must also be narrower than the spectral scale over which $\eta_\textrm{abs}$ changes appreciably. In general, continuous-wave operation is preferred as high peak fluence (\textit{i.e.}, pulsed) sources risk the onset of two-photon absorption. For semiconductors, undesired Auger carrier recombination is also avoided.\cite{klimov1997auger,klimov2000quantization,shen2018unexpectedly} 

More crucial is that the laser excitation wavelength, $\lambda_\textrm{exc}$, be greater than $\bar\lambda_\textrm{em}$ so that $P_\textrm{cool}>0$. A balance exists though. Too small a $\Delta E$ results in a small $P_\textrm{cool}$-value that is readily overwhelmed by any heating in the system. Too large a $\Delta E$-value results in insufficient cooling efficiencies/powers due to decreases in $\eta_\textrm{abs}$ (\textbf{Equation \ref{eqn:Pcool_equation_wavelength}}).

For RE\textsuperscript{3+}-doped materials, fixed wavelength sources are adequate for cooling studies. This is because so long as no phase transitions occur, the crystal-field energies of RE\textsuperscript{3+}-doped materials are largely temperature independent. To first order, crystal-field splittings increase as $R^{-5}$ with decreasing RE\textsuperscript{3+}-ligand bond length, $R$.\cite{song2022basic} In a YLF crystal, cooling from 300 to 100 K causes a lattice contraction of $\sim 0.187\%$ and $\sim 0.153\%$ along $a$ and $c$ axes, respectively.\cite{aggarwal2005measurement} Although the $R^{-5}$ relationship points to an increase of the crystal field splitting by $\sim 0.85\%$ in YLF:Yb\textsuperscript{3+}, the corresponding $E4 \rightarrow E5$ transition wavelength in \textbf{Figure \ref{fig:Figure_2}} only changes by a fraction of a nm, from 1019.26 nm (300 K) to 1019.60 nm (100 K).\cite{seletskiy2012cryogenic}. Such small changes make it possible to use fixed-wavelength lasers, tuned to the absorption transition of a given RE\textsuperscript{3+}-doped material, to conduct RE\textsuperscript{3+} cooling studies over a wide temperature range.
 
For semiconductors with relatively broad absorption band edges, detunings of a few tens of meV below the mean emission energy are typical. As illustration, in CdS nanobelts\cite{Zhang2013Laser} (CsPbBr$_{3}$ nanocrystals\cite{Roman2020Optically}), $\Delta E=38-119$ meV ($\Delta E=70-90$ meV). Unlike RE\textsuperscript{3+} systems, though, temperature-dependent absorption edges, as parameterized by material-specific Varshni relationships\cite{vurgaftman2001band,mannino2020temperature}, necessitate tunable laser sources to maintain (or vary) $\Delta E$ with changing specimen temperature.

\subsection{Appropriate non-contact thermometry}\label{subsect:thermometry}
Crucial to solid state laser cooling are reliable temperature measurement approaches for performance benchmarking and optimization. To this end, commercial/conventional temperature sensors have been successfully used to measure the temperature of a coldfinger\cite{hehlen2018}, fully shielded from the pump and PL (see \textbf{Section \ref{subsect:cryocooler}}). Miniaturized thermocouples, glued to the back of optical mirrors deposited onto ZBLAN:Yb\textsuperscript{3+} glasses, have also been used.\cite{thiede2005cooling} The approach, however, reduces $P_\textrm{cool}$ by an estimated 10-15\% due to photoluminescence absorption by unshielded parts of the electrical probe. These complications and accompanying performance limitations at cryogenic temperatures explain why non-contact approaches have since emerged as the preferred means of measuring temperature in solid-state optical cooling measurements.

Two general categories of non-contact thermometry are discussed in what follows. Passive methods assess temperature by analyzing information emitted by a cooling medium during laser cooling (either thermal radiation or ASPL). Active approaches require an auxiliary (probe) laser or other temperature reporter, in addition to the pump laser, to measure a cooling specimen's thermometric properties.

\subsubsection{Passive, non-contact thermometry}\label{subsubsect:passive_thermometry}
\textit{Thermal imaging.} A material being optically cooled emits infrared (blackbody) radiation that carries information about its average temperature. Emitted blackbody radiation can be monitored using calibrated thermal imagers such as microbolometers and focal-plane InSb, InAs, and HgCdTe arrays. While convenient, device temperature measurement ranges are often limited. Moreover, in some instances, data quality is adversely impacted by noise due to liquid cryogen boil-off. The low spectral emissivity of many semiconductors (\textit{e.g.}, GaAs) additionally restricts use of thermal imaging for their thermometry.

\textit{Differential Luminescence Thermometry (DLT).} PL emitted by specimens can simultaneously be used to ascertain cooling. This entails monitoring temperature-induced changes to emission lineshapes and energies. One approach, called differential luminescence thermometry, involves a differential analysis of PL spectra to monitor the temperature of RE\textsuperscript{3+}\cite{Mungan1997PRL,Patterson2007SPIE} and semiconductor\cite{imangholi2006differential} specimens during cooling. Because DLT spectral information is read out using charge-coupled device (CCD) or complementary-metal-oxide-semiconductor (CMOS) arrays, coupled to dispersive elements, among immediate limitations of the technique are its sensitivity and its speed. The former is determined by the spectral resolution of employed dispersive elements. The latter is established by CCD/CMOS readout rates. Among workarounds is the use of dispersive elements coupled to balanced photodiode pairs to monitor temperature-induced ASPL changes.\cite{seletskiy2011oe,patterson2010measurement}

At a deeper level, the temperature reporting fidelity of DLT is highly premised on the precise calibration of a specimen's PL lineshape as a function of temperature, observation angle, and optical mean-path length. This is because DLT is sensitive to reabsorption-induced changes to PL lineshapes that ultimately impact corresponding DLT traces. This may result in erroneous temperature measurements. The issue arises in large optical pathlength specimens or those that are optically dense. It also arises when multipass excitation schemes are used to maximize $P_\textrm{abs}$ in a cooling specimen.\cite{Kock2025UNM} Reabsorption-related issues are particularly problematic for simpler DLT variants that involve the ratiometric measurement of PL peaks.\cite{Bigotta2007SPIE} 

\textit{Other approaches.} Pump-induced ASPL can also be used to estimate temperature. This entails monitoring integrated ASPL intensities, taking advantage of the fact that $\eta_\textrm{c}$ is exquisitely sensitive to the temperature-dependence of $\eta_\textrm{abs}$. For CsPbBr\textsubscript{3} nanocrystals, a near exponential temperature dependence of the ASPL intensity has been used to establish specimen temperatures.\cite{zhang2020up,Roman2020Optically}

\subsubsection{Active, non-contact thermometry}\label{subsubsect:active_thermometry}
\textit{PL spectroscopy.}
Although uncommon, above gap-excited PL has been used to measure specimen temperatures. The approach has been used to estimate GaAs lattice temperatures during transient cooling measurements\cite{Lippmann2019CLEO} and has involved fitting the exponential decay of the high-energy tail in PL spectra to establish lineshape variations, stemming from thermal changes to phonon populations. Analogous PL thermometry on a YLF:Yb\textsuperscript{3+} crystal has been carried out using above gap-excited PL from a GaAs/GaInP heterostructure crystal connected to it.\cite{seletskiy2010oe}

\textit{Photothermal Deflection Spectroscopy (PDS).} Among active, non-contact thermometries that involve the introduction of a secondary probe laser, photothermal deflection spectroscopy is a commonly employed technique. It was, in fact, originally used to validate the first observation of solid-state optical refrigeration.\cite{epstein1995observation} 

In PDS, a (cooling) pump laser produces a radially-symmetric thermal lens in the illumination region. The lens then deflects an axially-shifted, probe beam that co-propagates through the specimen parallel to the pump.\cite{bowman2000new} Observed deflection angles are proportional to the sign of the local, radial thermal gradient. Deflection angles can therefore be used to discern cooling from heating. PDS requires exquisite alignment and understanding of the interplay between temperature-induced and population-induced lensing for accurate temperature measurements.\cite{Silva2013OL}

\textit{Optical interferometry.} Alternative, active non-contact techniques use optical interferometry to monitor temperature via temperature-induced changes to specimen refractive indices. An example is the homodyne detection of reflected probe light, which picks up a temperature-dependent phase shift upon transmission through the specimen, and a reference beam, which remains unaffected. The resulting interference signal possesses a magnitude proportional to $\Delta T$.\cite{hoyt2000PRL} While having the potential to offer extreme sensitivity, interferometric readout introduces measurement sensitivity to vibrations of requisite vacuum equipment. This introduces additional stabilization requirements that increase experimental complexity. A related approach uses fiber Bragg gratings (FBGs)\cite{Filho2014OE} directly glued to cooling specimens where the monolithic design of the interferometer makes it largely insensitive to environmental noise.\cite{Filho2013OE}

\textit{Cantilever approaches.} Finally, for very small cooling specimens (\textit{e.g.}, RE\textsuperscript{3+}-doped microcrystals \cite{dobretsova2021hydrothermal}), thermometric information can be extracted from temperature-induced changes to vibrational quality-factors of micrometer-sized cantilevers.\cite{pant2018optomechanical} In the approach, specimens are deposited atop cantilevers. Their subsequent cooling induces changes to the cantilever's mechanical response. A balanced photodiode then records the time domain response of scattered probe light off the cantilever's back whereupon the signal is Fourier transformed. Temperature information is then extracted from the linewidth of the obtained frequency response.

\subsubsection{Thermometry guidelines}\label{subsubsect:guidelines_thermometry}
Among above-described passive and active non-contact thermometries, thermal cameras remain the tool of choice for near room temperature experiments in RE\textsuperscript{3+}-based cooling studies. This includes its use to assess cooling efficiencies within the context of Laser-Induced Thermal Modulation Spectroscopy (LITMoS).\cite{Hoyt2003JOSAB}

At cryogenic temperatures, DLT remains the tool of choice. Of note then is recent work on sub-130 K and high-cooling-power operation in YLF:Yb\textsuperscript{3+}, which has revealed a sensitivity of DLT thermometry to the details of its signal collection. Namely, $P_\textrm{abs}$ is often maximized in cooling measurements by using a multipass pump scheme. What results, however, is a complex, three-dimensional network of passes through the cooling medium. Because DLT is sensitive to reabsorption effects, local variations of pump positions relative to the detection region impact DLT signals and can lead to erroneous conclusions.\cite{Kock2025UNM} 

Ongoing efforts have therefore worked to mitigate DLT's reabsorption sensitivity by performing a comprehensive, temperature-dependent calibration procedure that includes mapping PL lineshape dependencies with observation direction relative to the optical axis (if any) of the cooling medium and to the mean optical path ASPL takes to escape the specimen. The resulting calibration yields a laboratory- and setting-agnostic DLT methodology\cite{Kock2025UNM},  applicable to arbitrary specimens.

\section{Results}\label{section:results}
\subsection{RE\textsuperscript{3+}-doped material cooling milestones}\label{subsect:RE3_milestones}
In 1995, Epstein and co-workers at Los Alamos National Laboratory used a glassy ZBLANP (48\%ZrF$_4$-17\%BaF$_2$-4\%LaF$_3$-3\%AlF$_3$-20\%NaF-8\%PbF$_2$, mol\% \cite{poulin2024heavy}) fiber preform with a 1 wt\% Yb\textsuperscript{3+}-doped core to experimentally demonstrate net optical refrigeration of a solid for the first time.\cite{epstein1995observation} Cooling was induced by exciting the sample with a Ti:sapphire laser at $\lambda_\textrm{exc}=1020$ nm to achieve a $\Delta T\sim0.3$ K net cooling below room temperature. Temperature changes were probed using both PDS and thermal imaging. Subsequent improvements/increases in sample quality, absorbed excitation power, and thermal management resulted in cooling ZBLANP:Yb$^{3+}$ to a final temperature of $T_\textrm{final}\sim 208$ K from room temperature ($\lambda_\textrm{exc}=1026$ nm, $P_\textrm{exc}=9.6$ W).\cite{thiede2005cooling} \textbf{Table \ref{table:RE3+_glass_milestones}} summarizes additional Yb\textsuperscript{3+}-doped glass cooling milestones.

\begin{table}[h]
\begin{center}
\caption{Yb\textsuperscript{3+}-doped glass cooling milestones and key metrics.}
\begin{threeparttable}[t]
\begin{tabular}{ccccccccc}
\hline
Year & Host glass & Doping & $\Delta T$ (K) & $T_\textrm{final}$ (K) & Method & $\lambda_\textrm{exc}$ (nm) & $P_\textrm{exc}$ (W) & Reference\\
\hline
\hline
1995 & ZBLANP & 1 wt\% & 0.3 & $\textrm{RT}-\Delta T$ & PDS, IR & 1020 & 0.70 & \cite{epstein1995observation}\\
2002 & CNBZn & 1 mol\% & - & - & PDS & 988-1025 & - & \cite{fernandez2002origin}\\
2005 & ZBLANP & 2 wt\% & 87 & 208 & TC & 1026 & 9.6 & \cite{thiede2005cooling}\\
2006 & ABCYS & 2 wt\% & 0.13 & $\textrm{RT}-\Delta T$ & IR & 1020-1050 & - & \cite{guiheen2006yb3+}\\
2023 & GAYY & $4.0\times 10^{26}$ m$^{-3}$ & 0.9 & $\textrm{RT}-\Delta T$ & FBG, IR & 1029 & 6.5 & \cite{thomas2023demonstration}\\
2024 & SiO$_2$ & $6.56\times 10^{25}$ m$^{-3}$ & 67 & $\textrm{RT}-\Delta T$ & PL & 1032 & 97 & \cite{topper2024laser}\\
\hline
\end{tabular}
\begin{tablenotes}
\item[1] PDS=Photothermal deflection spectroscopy, IR=Thermal camera, TC=Thermocouple, FBG=Fiber Bragg grating, PL=Photoluminescence
\item[2] RT=Room temperature (exact temperature not specified)
\item[3] ZBLANP=48\%ZrF$_4$-17\%BaF$_2$-4\%LaF$_3$-3\%AlF$_3$-20\%NaF-8\%PbF$_2$ (mol\%)
\item[4] CNBZn=CdF$_2$-CdCl$_2$-NaF-BaF$_2$-BaCl$_2$-ZnF$_2$
\item[5] ABCYS=40\%AlF$_3$-12\%BaF$_2$-22\%CaF$_2$-16\%YF$_3$-10\%SrF$_2$ (mol\%)
\item[6] GAYY=SiO$_2$ codoped with Al$^{3+}$ (7.6 at.\%), Y$^{3+}$ (4.6 at.\%), and Yb$^{3+}$ 
\end{tablenotes}
\end{threeparttable}
\label{table:RE3+_glass_milestones}
\end{center}
\end{table}

In 2000, Bowman and Mungan demonstrated the first cooling of a RE\textsuperscript{3+}-doped crystal. This entailed using PDS to observe net cooling in a KGd(WO\textsubscript{4})\textsubscript{2}:3.5\%Yb\textsuperscript{3+} crystal (KGW:Yb\textsuperscript{3+}, $\lambda_\textrm{exc}=1008–1026$ nm).\cite{bowman2000new} In 2001, Epstein \textit{et al.} cooled a Y$_3$Al$_5$O$_{12}$:2.3\%Yb$^{3+}$ (YAG:Yb$^{3+}$) crystal by $\Delta T=8.9$ K from room temperature ($\lambda_\textrm{exc}=1030$ nm, $P_\textrm{exc}=1.8$ W) and a Y$_2$SiO$_5$:5\%Yb$^{3+}$ (YSO:Yb$^{3+}$) crystal by $\Delta T\sim1$ K ($\lambda_\textrm{exc}=1050$ nm).\cite{epstein2001measurements} Subsequent work by Bigotta \textit{et al.} in 2007 demonstrated promising results by cooling a YLiF$_4$:5\%Yb$^{3+}$ (YLF:Yb$^{3+}$) crystal by $\Delta T=69$ K from room temperature ($\lambda_\textrm{exc}=1030$ nm, $P_\textrm{exc}=15$ W).\cite{bigotta2007laser} 

Seletskiy \textit{et al.} subsequently modeled the cooling performance of YLF:Yb\textsuperscript{3+} by taking into account its temperature-dependent absorption. They predicted the possibility of cooling YLF:Yb\textsuperscript{3+} to 110 K when pumped at $\lambda_\textrm{exc}=1020$ nm to be in resonance with Yb\textsuperscript{3+}'s $E4 \rightarrow E5$ crystal-field transition (\textbf{Figure \ref{fig:Figure_2}}).\cite{seletskiy2010laser} Seletskiy \textit{et al.} went on to demonstrate YLF:Yb\textsuperscript{3+} cooling to $T_\textrm{final}\sim 155$ K from room temperature [$\eta_\textrm{EQE}=0.995\pm 0.001$, $\alpha_\textrm{b}=(4.2\pm 0.2)\times 10^{-4}$ cm$^{-1}$] using $\lambda_{\textrm{exc}}$ = 1023 nm. These results pointed to suppressed inhomogeneous broadening in crystalline, as opposed to glassy, hosts offering significant enhancements of $\eta_\textrm{abs}$, thus opening up an era of cryogenic research in solid-state laser refrigeration. 

The first verification of solid-state optical refrigeration to cryogenic temperatures came in 2013 when Melgaard \textit{et al.} cooled a YLF:5\%Yb$^{3+}$ crystal [$\eta_\textrm{EQE}=0.994\pm 0.001$, $\alpha_\textrm{b}=(4.4\pm 0.2)\times 10^{-4}$ cm$^{-1}$], placed inside a Herriott cell, to $T_\textrm{final}=123.7\pm1.0$ K from room temperature ($\lambda_\textrm{exc}=1020.7$ nm, $P_\textrm{exc}=45$ W, $P_\textrm{abs,tot}=18$ W).\cite{melgaard2013optical} The crystal was further cooled to $T_\textrm{final}=118.7\pm1$ K by lowering the temperature of a clamshell surrounding the crystal to 208 K. Further improvements in sample purity [$\eta_\textrm{EQE}=0.996\pm 0.001$, $\alpha_\textrm{b}\leq(1 \pm 0.1)\times10^{-4}$ cm\textsuperscript{-1}] and optical refrigerator design improved the cooling to $T_\textrm{final}=91$ K in a YLF:10\%Yb\textsuperscript{3+} crystal.\cite{melgaard2016solid} YLF:5\%Yb$^{3+}$ was subsequently cooled to $T_\textrm{final}=87$ K from room temperature by using an astigmatic Herriott cell to increase the number of pump laser passes through the sample from 22 \cite{melgaard2016solid} to $>300$ while simultaneously increasing the specimen's absorptance to nearly 100\%.\cite{gragossian2017optical} 

To date, the lowest $T_\textrm{final}$ reported for YLF:Yb$^{3+}$ is by Volpi \textit{et al.} who cooled very lightly Tm$^{3+}$ co-doped ($\sim 16$ ppm) YLF:5\%Yb$^{3+}$ \cite{di2014influence} to $T_\textrm{final}=87$ K from room temperature ($\lambda_\textrm{exc}=1020$ nm, $P_\textrm{exc}=50$ W).\cite{volpi2019optical} The reported $T_\textrm{final}=87$ K is lower than a calculated YLF:Yb$^{3+}$ gMAT of 110-120 K at 1020 nm, based on temperature-independent $\eta_\textrm{EQE}=0.994$ and $\alpha_\textrm{b}=2\times10^{-4}$ cm$^{-1}$ values. This suggests a temperature-dependent $\alpha_\textrm{b}$-value in YLF:Yb$^{3+}$, which decreases with decreasing temperature. In support of this, a direct measurement has revealed an order of magnitude decrease in $\alpha_\textrm{b}$ between 300 K and 100 K. A revised YLF:Yb$^{3+}$ gMAT is therefore $\sim 70$ K.\cite{volpi2019optical} \textbf{Table \ref{table:RE3+_crystal_milestones}} summarizes important Yb$^{3+}$-doped crystal cooling results.

\begin{table}[h]
\begin{center}
\caption{Yb\textsuperscript{3+}-doped crystal cooling milestones.}
\begin{threeparttable}[t]
\scriptsize
\begin{tabular}{ccccccccc}
\hline
Year & Host crystal & Doping\tnote{1} & $\Delta T$ (K) & $T_\textrm{final}$ (K) & Method\tnote{2} & $\lambda_\textrm{exc}$ (nm) & $P_\textrm{exc}$ (W) & Reference\\
\hline
\hline
2000 & KGd(WO\textsubscript{4})\textsubscript{2} & 3.5 at.\% & - & - & PDS & 1000, 1028 & - & \cite{bowman2000new}\\
2001 & YAG & 2.3 at.\% & 8.9 & - & IR & 1030 & 0.75 & \cite{epstein2001measurements}\\
2001 & Y\textsubscript{2}SiO\textsubscript{5} & 5 at.\% & 1 & - & IR & 1050 & - & \cite{epstein2001measurements}\\
2002 & KPb\textsubscript{2}Cl\textsubscript{5} & $\sim$5$\times10^{19} \ \mathrm{cm}^{-3}$ & - & - & PDS & 985 - 1035 & - & \cite{fernandez2002origin, mendioroz2002anti}\\
2006,2007 & BaY\textsubscript{2}F\textsubscript{8} & 2.5 at.\% & 4 & - & IR & 1025 & 3 & \cite{bigotta2007laser,bigotta2006spectroscopic,bigotta2006laser}\\
2007 & YLF & 5 at.\% & 69 & - & PL & 1030 & 15 & \cite{bigotta2007laser}\\
2013 & YLF & 5 at.\% & 162/89\tnote{3} & 124/119\tnote{3} & PL & 1020 & 45 & \cite{melgaard2013optical}\\
2016 & YLF & 10 at.\% & 178.5 & 91 & PL & 1020 & 54 & \cite{melgaard2016solid}\\
2017 & YLF & 5 at.\% & $\sim$200 & 87 & PL & 1020 & - & \cite{gragossian2017optical, volpi2019optical}\\
2018 & KYF\textsubscript{4} & 10 at.\% & 1.8 & - & IR & 1025 & 1.5 & \cite{volpi2018anti}\\
2020 & KY(WO\textsubscript{4})\textsubscript{2} & 1 at.\% & 1.5 & - & IR & 1025 & - & \cite{cheng2020laser} \\
2021 & YAG & 3 at.\% & 81 & 219 & PL & 1030 & 37 & \cite{zhong2021optical}\\
2022 & LuLiF\textsubscript{4} & 7.5 at.\% & 145 & 121 & PL & 1020 & 80 & \cite{lei2022laser} \\
2022 & CaF\textsubscript{2} & 5 at.\% & - & - & IR & 1019 - 1080 & - & \cite{puschel2022solid} \\
2022 & SrF\textsubscript{2} & 5 at.\% & - & - & IR & 1012 - 1075 & - & \cite{puschel2022solid} \\
2022,2023 & KY\textsubscript{3}F\textsubscript{10} & 3 at.\% & - & - & IR & 1002 - 1067 & - & \cite{puschel2022laser, puschel2023ytterbium} \\
2024 & CaF\textsubscript{2} & 5 at.\% & 97 & 200 & PL & 1020 & 100 & \cite{puschel2024thesis} \\
2024 & BaY\textsubscript{2}F\textsubscript{8} & 5 at.\% & 100 & 194 & PL & 1020 & 100 & \cite{puschel2024thesis} \\
2024 & KY\textsubscript{3}F\textsubscript{10} & 3 at.\% & 120 & 177 & PL & 1020 & 100 & \cite{puschel2024thesis} \\
\hline
\end{tabular}
\begin{tablenotes}
\item[1] Doping concentration in atomic percentage (at.\%) refers to the ratio with respect to the cations substituted by Yb\textsuperscript{3+}. For example, the chemical formula corresponding to YLF:5 at.\% Yb\textsuperscript{3+} is (Yb\textsubscript{0.05}Y\textsubscript{0.95})LiF\textsubscript{4}.
\item[2] PDS=Photothermal deflection spectroscopy, IR=Thermal camera, PL=Photoluminescence
\item[3] Results with the temperature of the crystal's surrounding at 285/208 K.
\end{tablenotes}
\end{threeparttable}
\label{table:RE3+_crystal_milestones}
\end{center}
\end{table}

Above efforts along with a low gMAT have made YLF:Yb$^{3+}$ the workhorse material for high-power cooling experiments seeking to realize solid-state, cryogenic optical refrigeration. This has been aided by progressive improvements in the quality of Czochralski-grown YLF:Yb$^{3+}$ over the last decade. Today, crystals with $\eta_\textrm{EQE}>0.99$ and $\alpha_\textrm{b}\approx 10^{-4}$ cm\textsuperscript{-1} are routinely reported.

To achieve lower temperatures with YLF:Yb$^{3+}$, one must increase Yb$^{3+}$'s concentration while lowering $\alpha_\textrm{b}$ by reducing impurity concentrations.\cite{puschel2021temperature} This requires more stringent precursor purification processes, \textit{e.g.}, by zone refining YF$_3$, LiF, and YbF$_3$ precursors and/or polycrystalline YLF:Yb$^{3+}$ prior to single crystal growth. Alternatively, increased cooling performance at cryogenic temperatures is possible by choosing a different Yb$^{3+}$ host that possesses improved spectroscopic properties. Here, Püschel \textit{et al.} have demonstrated cooling KY$_3$F$_{10}$:Yb\textsuperscript{3+} (a system that possesses a laser cooling figure-of-merit more than five times greater than that of YLF:Yb$^{3+}$ at temperatures below 200 K\cite{puschel2022laser}) by $\Delta T=177$ K from room temperature ($\lambda_\textrm{exc}=1020$ nm, $P_\textrm{exc}=100$ W).\cite{puschel2024thesis} This promising development suggests future opportunities for improving the performance of Yb$^{3+}$-doped fluoride crystals. 

\subsubsection{Beyond Yb\textsuperscript{3+}}\label{subsubsect:beyond_Yb3}
Given that \textbf{Equation \ref{eqn:eta_cool_equation_wavelength}} has shown $\eta_\textrm{c}$ to scale as ${\Delta E\over(\bar{E}_\textrm{em}-\Delta E)}$ (in the limit that $\eta_\textrm{em}$$\eta_\textrm{abs}$ = 1), using cooling ions with $\bar E_\textrm{em}$ lower than that of Yb\textsuperscript{3+} (\textbf{Table \ref{table:RE3+_summary}}) should increase $\eta_\textrm{c}$ and enable greater $P_\textrm{cool}$-values for the same $P_\textrm{abs}$. Several studies have therefore pursued this optimization strategy to demonstrate laser cooling with Er\textsuperscript{3+}, Tm\textsuperscript{3+}, Ho\textsuperscript{3+}, and Cr\textsuperscript{3+}. For these ions, $m>8$ for all halide hosts. Large $\eta_\textrm{ext}$ are therefore possible. \textbf{Table \ref{table:beyond_yb3+}} summarizes these studies.

\begin{table}[h]
\begin{center}
\caption{RE\textsuperscript{3+}-doped crystal/glass cooling beyond Yb\textsuperscript{3+}.}
\begin{threeparttable}[t]
\begin{tabular}{cccccc}
\hline
Host & Doping\% & $\Delta T$ (K) & $\lambda_\textrm{exc}$ (nm) & $P_\textrm{exc}$ (W) & Reference\\
\hline
\hline
\multicolumn{6}{c}{Er\textsuperscript{3+}}\\
CNBZn\tnote{1}\color{black} & 0.5 mol\%\tnote{2} & 0.5$\pm$0.1 & 860 & 1.9 & \cite{fernandez2006anti}\\
KPb\textsubscript{2}Cl\textsubscript{5} & 0.5 mol\%\tnote{3} & 0.7$\pm$0.1 & 879 & 1.9 & \cite{fernandez2006anti, condon2009optical}\\
\hline
\multicolumn{6}{c}{Tm\textsuperscript{3+}}\\
ZBLANP & 1 wt.\% & 24 & 1900 & 4.5 & \cite{hoyt2000PRL,hoyt2003laser}\\
BaY\textsubscript{2}F\textsubscript{8} & 1.2 at.\% & 3.2 & 1855 & 4.4 & \cite{patterson2008anti}\\
YLF & 1 at.\% / 2 at.\% & - & 1860-1945/1880-2000 & - & \cite{rostami2019tm}\\
InF & 1 mol\% & 2.4 & 1900 & 4.4 & \cite{nobel2025optical}\\
\hline
\multicolumn{6}{c}{Ho\textsuperscript{3+}}\\
YLF & 1 at.\% & - & 2056-2220 & - & \cite{rostami2019observation}\\
BaY\textsubscript{2}F\textsubscript{8} & 1 at.\% & - & 2023-2100 & - & \cite{rostami2021observation}\\
\hline
\multicolumn{6}{c}{Cr\textsuperscript{3+}}\\
LiSAF & 1 at.\% & - & 1000 & - & \cite{silva2025observation}\\
\hline
\end{tabular}
\begin{tablenotes}
\item[1] CNBZn = CdF\textsubscript{2}-CdCl\textsubscript{2}-NaF-BaF\textsubscript{2}-BaCl\textsubscript{2}-ZnF\textsubscript{2}.
\item[2] 0.5 mol\% of ErCl\textsubscript{3} doping.
\item[3] 0.5 mol\% of ErF\textsubscript{3} doping.
\end{tablenotes}
\end{threeparttable}
\label{table:beyond_yb3+}
\end{center}
\end{table}

The next ion in this series of decreasing $\bar{E}_\textrm{em}$ and potentially increasing $\eta_\textrm{c}$ is Dy\textsuperscript{3+}. With $\bar{E}_\textrm{em}\approx 3200$ cm\textsuperscript{-1} (\textbf{Table \ref{table:RE3+_summary}}), host materials with $\hbar\omega_\textrm{max}<400$ cm\textsuperscript{-1} are required to achieve high $\eta_\textrm{EQE}$-values. This points to the use of chloride, bromide, or iodide hosts. Notably, solid-state laser cooling of Dy\textsuperscript{3+} in any host remains to be demonstrated.

\subsubsection{Yb\textsuperscript{3+}-doped nanostructures and other morphologies}\label{subsubsect:yb3_nanostructures}
In tandem, significant work has focused on demonstrating solid-state laser cooling in RE\textsuperscript{3+}-doped particles.\cite{nemova2022laser} Among examples, levitated YLF:10\%Yb\textsuperscript{3+} nanocrystals have been refrigerated to $T_\textrm{final}=130$ K by pumping them at $\lambda_\textrm{exc}=1020$ nm in a single-beam dipole trap.\cite{rahman2017laser} Such particles have also been cooled by $\Delta T=10$ K in a D\textsubscript{2}O medium\cite{roder2015laser} and by $\Delta T=23.6$ K when placed atop a CdS cantilever.\cite{pant2020solid} Similar results were achieved with LLF:10\%Yb\textsuperscript{3+} microcrystals, which were cooled by $\Delta T=20.4$ K in vacuum\cite{dobretsova2021hydrothermal} and by $\Delta T=5$ K in deionized water.\cite{dobretsova2022safe}

Cooling has also been investigated in the potassium lutetium fluorides (K\textsubscript{2}LuF\textsubscript{5}, KLuF\textsubscript{4}, KLu\textsubscript{2}F\textsubscript{7}), in KLu\textsubscript{3}F\textsubscript{10} nanocrystals \cite{xia2021hydrothermal}, and in $\beta$-NaYF\textsubscript{4}:10\%Yb\textsuperscript{3+} nanowires. The latter were cooled by $\Delta T>9$ K in a D\textsubscript{2}O medium ($\lambda_\textrm{exc}=1020$ nm)\cite{zhou2016laser, ortiz2021laser} while optically levitated nanocrystals of the same material were cooled by $\Delta T=42$ K ($\lambda_\textrm{exc}=1020$ nm).\cite{luntz2021laser} Laser cooling has also been observed in porous $\alpha$-NaYF\textsubscript{4} nanocubes\cite{forberger2024synthesis} and CaF\textsubscript{2}:Yb\textsuperscript{3+} spherical microcrystals.\cite{chen2024optical}

Finally, laser cooling has been demonstrated in optical fibers. A ZBLAN:Yb\textsuperscript{3+} fiber has been cooled by $\Delta T>7$ K ($\lambda_\textrm{exc}=1030$ nm, $P_\textrm{exc}=36.5$ W)\cite{xie2023laser} while a Yb\textsuperscript{3+}-doped silica fiber has been cooled by $\Delta T=18.4$ K ($\lambda_\textrm{exc}=1035$ nm, $P_\textrm{exc}=20$ W).\cite{topper2021laser}

\subsection{Semiconductor laser cooling claims}\label{subsect:semiconductor_cooling_claims}
Early semiconductor cooling attempts focused on GaAs quantum wells and  GaAs/GaInP heterostructures. This stemmed from the availability of high quality specimens grown using molecular beam epitaxy.\cite{Gauck1997_GaAsGaInP, Bender2013_GaAsGaInP,giannini2021near,finkeissen1999cooling} Although Sheik-Bahae and co-workers realized a record setting GaAs/GaInP double heterostructure $\eta_\textrm{EQE}$ of $0.995\pm 0.001$ (a value above $\eta_\textrm{EQE,crit}=0.987$), no cooling was observed. This was ultimately attributed to too low a $\eta_\textrm{abs}$-value ($\eta_\textrm{abs}\sim 0.901$, $\alpha_\textrm{b}$ = 91 cm${^{-1}}$, \textbf{Table \ref{table:semiconductors}}) due to the presence of a parasitic background absorption that induced heating.\cite{Gauck1997_GaAsGaInP,Bender2013_GaAsGaInP} While the origin of this background absorption remains unknown, it is thought to relate to point defects at GaAs/GaInP heterojunctions.

The one GaAs laser cooling claim to emerge from these early studies\cite{finkeissen1999cooling} suggested a $\Delta T\sim 7$ K cooling below liquid nitrogen temperature. $\Delta T$ was inferred by changes to the intensity ratio of heavy hole to light hole emission. This claim, though, was refuted by Sheik-Bahae and co-workers\cite{hasselbeck2007effect}, who attributed observed spectral changes to photocarrier density–dependent screening effects that altered exciton binding energies.

More recent reports have investigated laser cooling in semiconductor nanostructures. In 2013, Xiong and co-workers reported cooling a chemical vapor deposition-grown CdS nanobelt by $\Delta T=40$ K from room temperature.\cite{Zhang2013Laser} Cooling was reported as a blueshift of the CdS emission when nanobelts were excited below gap. This contrasts to redshifts seen when exciting above gap. Despite these observations being in qualitative agreement with cooling, the claim has been challenged due to significant disparities in observed versus predicted cooling/heating timescales.\cite{morozov2019can} Up to a nine orders of magnitude difference in timescales exists. Moreover, there are concerns regarding the use of estimated, not measured, $\eta_\textrm{EQE}$-values and the lack of $\eta_\textrm{abs}$ measurements or estimates. This is especially important given the likely involvement of donor-acceptor states in the CdS nanobelt PL up-conversion process.\cite{xu2011dynamics, morozov2017defect}

Xiong and co-workers have also reported cooling lead-based, hybrid perovskite crystals, namely methylammonium lead iodide (MAPbI$_3$) microcrystals by $\Delta T=23$ K and two-dimensional (2D) Ruddlesden-Popper, phenethylammonium lead iodide (PhEPbI$_4$) crystals by $\Delta T=58.7$ K, both from room temperature\color{black}.\cite{ha2016laser} As with their earlier CdS nanobelt report, similar concerns exist over underreported cooling metrics and, more importantly, about observed cooling/heating timescales being significantly longer than predicted.

In the realm of colloidal quantum dots, there are three primary cooling claims for CdSe-based core/shell nanocrystals. The first is by Fontenot \textit{et al.} who report laser cooling CdSe/ZnS quantum dots by $\Delta T=2.3$ K from room temperature.\cite{fontenot2017optical,fontenot2016measuring} The second is by Decca \textit{et al.} who report a $\Delta T=0.37 \pm 0.08$~K cooling from room temperature of analogous CdSe/CdS quantum dots.\cite{Hua2020Optical} Finally, Peng and co-workers report a 0.18 K relative cooling of their CdSe/CdS quantum dots.\cite{Ye2021Phonon} The relative descriptor indicates suppressed heating.

For all-inorganic, CsPbBr$_3$ perovskite nanocrystals, Sheldon  \textit{et al.} reported a $\Delta T=25$ K cooling of their particles from room temperature.\cite{Roman2020Optically} Yamada \textit{et al.} have likewise reported a $\Delta T=9$ K cooling of CsPbBr$_3$ nanocrystal inclusions in zero dimensional Cs$_4$PbBr$_6$ from room temperature.\cite{Yamada2024Optical}

Despite all abovementioned results, stated claims of semiconductor nanostructure optical refrigeration remain highly debated. This stems from concerns about material quality and specifically as to whether said cooling materials possess requisite unity/near-unity $\eta_\textrm{EQE}$-values. Attesting to this, $\eta_\textrm{EQE}$-values have often been assumed or estimated. Additional concerns emerge from whether reported data adhere to known thermodynamic constraints, which dictate the extent of cooling (\textit{i.e.}, $T_\textrm{final}$) as well as corresponding cooling/heating timescales. Reference \cite{morozov2019can} provides more details about these concerns with existing semiconductor cooling claims. To standardize cooling assessments and to ensure their reliability going forward, we have therefore developed an experimental checklist that authors should follow when reporting semiconductor cooling claims, published in Reference \cite{Zhang2025Principles} and summarized in \textbf{Section \ref{subsect:reporting_standards}}.

\section{Applications}\label{section:applications}

\subsection{Optical cryocoolers}\label{subsect:cryocooler}
The low temperatures achieved with RE\textsuperscript{3+}-doped crystals position them as excellent candidates for creating vibration-free optical cryocoolers. Such a cooler is designed to refrigerate an arbitrary payload, typically by attaching it to a cold finger. The cold finger is connected to the cooling crystal via a thermal link, which provides good thermal contact. The entire system resides in vacuum to minimize convective heat loads. As such, a crycooler is a composite system, consisting of three subsystems: (I) a cooling sample at temperature $T_{\rm sample}$, (II) a thermal link at temperature $T_{\rm link}$, and (III) a cold finger/payload at temperature $T_{\rm payload}$.

To maximize the system's cooling power, the interplay of all thermal loads on the system is modeled. For the sample, a governing equation is\cite{seletskiy2010oe,Melgaard2011UNM,Meng2020UNM}
\begin{equation}
C(T)\frac{dT_{\rm sample}}{dt}=-{\dot{Q}}_\textrm{cool}+{\dot{Q}}_\textrm{env},
\label{eqn:T_change}
\end{equation}
where $C(T)=\rho c_\textrm{v}(T)V_\textrm{sample}$ is the sample's heat capacity given a density $\rho$, a specific heat capacity $c_\textrm{v}(T)$, and a sample volume $V_\textrm{sample}$. ${\dot{Q}}_{\rm env}$ is an environmental heat load and is
\begin{equation}
{\dot{Q}}_\textrm{env}=\frac{\kappa_\textrm{link}(T_{\rm l})A_\textrm{link}}{d_\textrm{l}}(T_\textrm{link}-T_\textrm{sample})+\frac{\epsilon_\textrm{sample}A_\textrm{sample}}{1+\chi}\sigma(T_\textrm{clamshell}^{4}-T_{\rm sample}^4),
\label{eqn:heat_load}
\end{equation}
where $\sigma$ is the Stefan-Boltzmann constant and the first term is a conductive heat load on the sample from the thermal link while the second is a radiative load. For brevity, a term describing a conductive heat load from sample mechanical supports (which can be minimized, see below) is omitted. The thermal link is parameterized by a cross-sectional area, $A_\textrm{link}$, an average length, $d_\textrm{link}$, and a thermal conductivity, $\kappa_\textrm{link}$ so that in the second term,
\begin{equation}
\chi=(1-\epsilon_\textrm{clamshell})\frac{\epsilon_\textrm{sample}A_\textrm{sample}}{\epsilon_\textrm{clamshell}A_\textrm{clamshell}},
\label{eqn:chi_radiative}
\end{equation}
where $A_\textrm{sample}$ and $A_\textrm{clamshell}$ are the surface areas of the sample and a surrounding enclosure (called a clamshell), each with thermal emissivity $\epsilon_\textrm{sample}$ and $\epsilon_\textrm{clamshell}$, respectively. 

Equations of state for subsystems II and III exclude ${\dot{Q}}_\textrm{cool}$ and follow the framework outlined in \textbf{Equations \ref{eqn:T_change}}-\textbf{\ref{eqn:chi_radiative}}, adapted to their respective environments (for details, see \cite{Meng2020UNM}). Temperatures $\{T_{\rm sample},T_{\rm link},T_{\rm payload}\}$ are obtained from a self-consistent solution to the coupled equations of state for the three subsystems. 

We now discuss how each subsystem is optimized to ensure maximal cooling power delivery to the payload. In short, the general strategy entails maximizing $\dot Q_{\rm cool}$, maximizing its delivery to the payload all while minimizing heat loads across subsystems. The payload must simultaneously be protected against absorption of ASPL and/or scattered pump light, either of which, if absorbed, could reduce or even negate available cooling power.

\textit{Sample design.}
To achieve useful optical cryocooler performance, cooling materials with a sufficiently low gMAT must be used. $P_\textrm{cool}=\eta_\textrm{c}P_\textrm{abs}$ (\textbf{Equation \ref{eqn:Pcool_equation_wavelength}}) is maximized by using a multi-pass pumping scheme, \textit{e.g.}, through use of an astigmatic Herriott cell.\cite{Gragossian2016OE} See \textbf{Sections \ref{section:experimentation}} and \textbf{\ref{section:results}} for additional considerations.

\textit{Thermal link design.}
The main function of the thermal link is to provide good thermal contact between subsystems I (sample) and III (payload). Unless the payload is transparent to the ASPL from the cooling sample (\textit{e.g.}, cooling a GaAs device with YLF:Yb\textsuperscript{3+} \cite{seletskiy2010oe}), the thermal link must simultaneously prevent emission reaching the payload. This is a critical issue, since the cooling power is only a fraction of the incoming laser power ($P_{\textrm{cool}} = \eta_{\textrm{c}}P_{\textrm{abs}}\approx \eta_{\textrm{c}}P_{\textrm{em}}$), which becomes small when $T_{\rm sample} \approx T_{\rm gMAT}$. Yet, the payload can potentially see the full magnitude of $P_{\rm em}$. Hence, payload optical isolation must considerably exceed (1-$\eta_{\rm c}$). In practice, this  amounts to a 99.5\% or higher requirement for stray light suppression.

\begin{figure}[htb]
\centering
\includegraphics[width=0.8\linewidth]{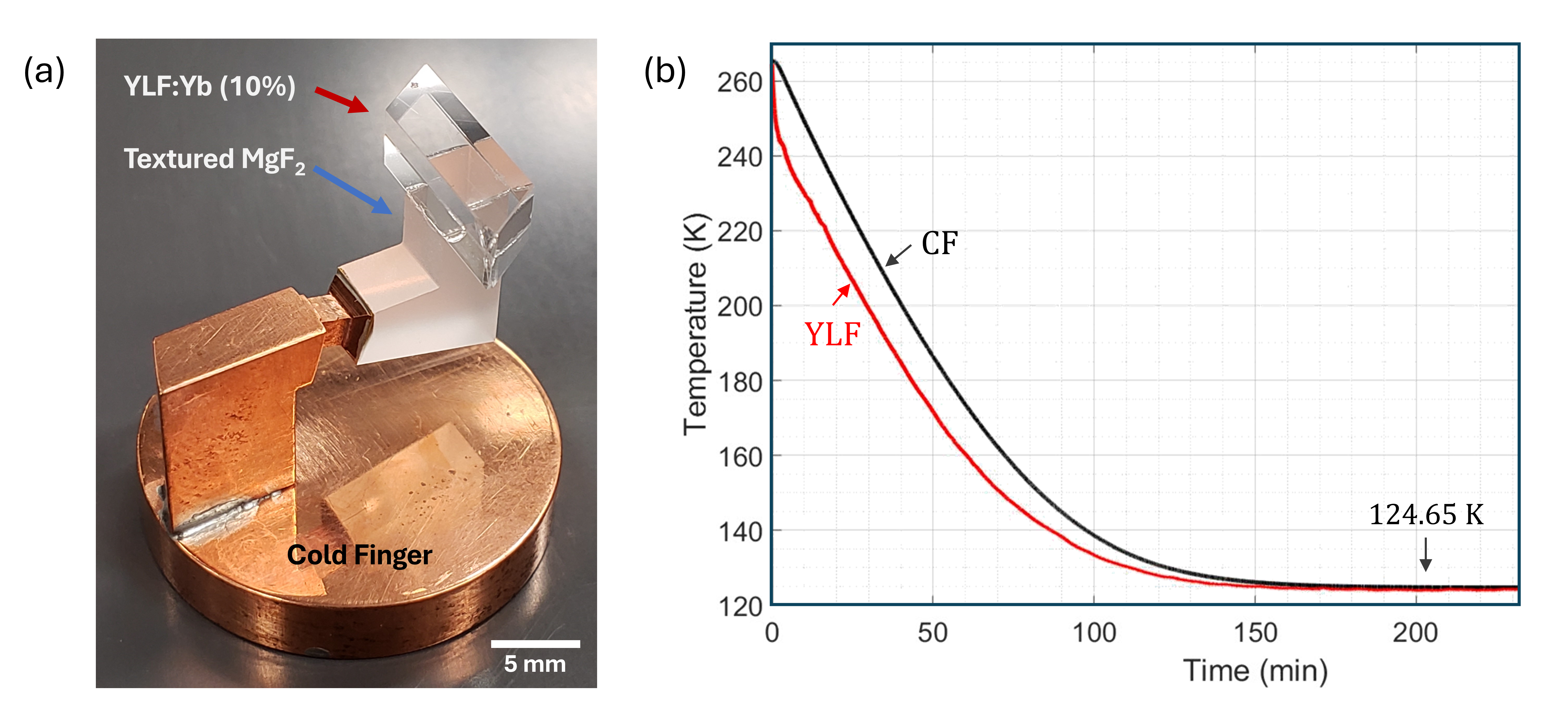}
\caption{(a) Photograph of a cryocooler. The cooling crystal (YLF:Yb\textsuperscript{3+} 10\%) is attached to a copper cold finger (CF) via a textured MgF\textsubscript{2} thermal link. Note that the clamshell has been removed for this photograph. (b) Measured temperature of YLF (via DLT) and CF (via silicon diode) during payload cooling to 125 K.\cite{kock2022ol}} 
\label{fig:Figure_3}
\end{figure}

Presently, the best thermal link designs are based on transparent materials, which fulfill criteria such as low parasitic absorption, high thermal conductivity, and a closely matched coefficient of thermal expansion (CTE) between cooling crystal and link.\cite{Meng2020UNM} In the case of a YLF:Yb\textsuperscript{3+} cooler, thermal link candidate materials include sapphire (suffers from a CTE mismatch), undoped YLF (low thermal conductivity), and MgF$_2$. To optically isolate payloads, one or more near-90-degree bends can be employed. An optimized geometry is highlighted in Reference \cite{hehlen2018}. Textured surfaces can also be used to limit ASPL waveguiding towards payloads (\textbf{Figure \ref{fig:Figure_3}a}). Ideally, adhesive-free bonding should be used to connect the cooling crystal to the thermal link \cite{hehlen2018}, although a very thin layer of cyanoacrylate adhesive has been successfully demonstrated.\cite{kock2022ol}

\textit{Conductive load minimization.}
Cryocooler mechanical supports typically dominate parasitic conductive heat loads. To minimize this (\textbf{Equation \ref{eqn:heat_load}}), low thermal conductivity supports such as a silica aerogel\cite{Fesmire_2019} or small cross-section and long length supports, like optical fibers or a Kevlar suspension\cite{TIMBIE1990271} can be implemented.

\textit{Radiative load minimization.}
A tight-fitting clamshell with surface area $A_{\rm clamshell}$ surrounds most of the cryocooler assembly (surface area, $A_{\rm system}$). This reduces the system's radiative load by maximizing $\chi$ in \textbf{Equation \ref{eqn:chi_radiative}} (\textit{i.e.}, $A_{\rm clamshell}\rightarrow A_{\rm system}$). Cooling sample and thermal link material properties determine their emissivities. The parasitic radiative load on the cold finger and any metallic component of the payload is reduced by mechanical polishing or surface gold-plating, lowering their respective thermal emissivities. The clamshell is coated with a solar-selective coating\cite{thiede2005cooling} that exhibits low thermal emissivity near thermal wavelengths of 10 $\mu$m. For Yb\textsuperscript{3+} cryocoolers, the coating is highly absorptive in the 1 $\mu$m wavelength range where Yb\textsuperscript{3+} fluoresces to suppresses scattered or reflected PL. The clamshell's close proximity to the cooler assembly (approximately 1 mm) along with additional baffling prevents direct or scattered ASPL or pump laser light from reaching the cold finger or payload.

In this way, a HgCdTe detector payload has been cooled to 135 K by a YLF:Yb\textsuperscript{3+} cryocooler.\cite{hehlen2018} More recent demonstrations have shown payload cooling to 125 K\cite{kock2022ol} (\textbf{Figure \ref{fig:Figure_3}b}), opening up opportunities for high-precision metrology applications.\cite{Kessler2012NaPho}

\subsection{Radiation-balanced lasers and amplifiers}\label{subsect:rbl_and_rba}
In 1999, Bowman established conditions where the heat produced by a laser's operation could be balanced by optical refrigeration to create an ``athermal'', radiation-balanced laser (RBL).\cite{Bowman1999JQE} Motivating this was the desire to overcome heating effects such as temperature-induced and spatially-varying birefringence. Furthermore, an athermal RBL would overcome power scaling roll-over issues that plague conventional high-power lasers, even in thin-disk configurations.\cite{Giesen1994}

A complementarity between RBL operation and optical refrigeration is seen through the lens of a heat engine. Shortly after the invention of the maser, Scovil and Schulz-DuBois established its formal connection to the Second Law of thermodynamics.\cite{Scovil1956PRL} These arguments were later extended to the laser.\cite{Kosloff2014ARPC} By defining a maser/laser efficiency, $\eta_\textrm{laser}={\nu_\textrm{l}\over\nu_\textrm{p}}$, where $\nu_\textrm{l}$ is the laser output frequency and $\nu_\textrm{p}$ is the pump frequency, Scovil and Schulz-DuBois demonstrated that $\eta_\textrm{laser}$ obeyed the Carnot limit wherein $\eta_\textrm{laser}\le \eta_\textrm{Carnot}=1-{T_\textrm{c}\over T_\textrm{h}}$. 

A three-level laser is therefore a heat engine that operates between a hot bath at a temperature, $T_\textrm{h}$, and a cold bath at a temperature, $T_\textrm{c}$, generating coherent radiation (\textit{i.e.}, work) at a frequency $\nu_\textrm{l}$. The hot reservoir provides energy, a portion of which becomes work. The remainder, in the form of entropy, is deposited into the cold reservoir. For a laser, the cold reservoir entails lattice phonons, spontaneous emission, parasitic non-radiative decay, and any other decay channel that does not produce work. Energy deposited into the cold reservoir irreversibly increases the laser's entropy via heating. 

When run in reverse, the laser heat engine acts as a heat pump, transferring heat from the cold reservoir to the hot one through inputted work. Optical refrigeration can therefore be visualized as a laser heat pump, which uses inputted, coherent laser radiation to pump heat from a cold reservoir (the material-to-be-cooled) to a hot one (photoluminescence emitted into the surroundings). The reversing of the laser heat engine should not be confused with recent demonstrations of ``time-reversed lasers'' where resonant
cavities are used to achieve a ``coherent perfect absorber''.\cite{Chong2010PRL} A true thermodynamic analogue requires that the intracavity element be cooled and has only been demonstrated through cavity-enhanced optical refrigeration.\cite{seletskiy2010APL,seletskiy2016laser}

Various lasing materials have been evaluated for their RBL suitability.\cite{bowman2000new,Nemova2021ApSci} The first RBL was demonstrated in 2002 using a bulk KGd(WO$_4$)$_2$:Yb\textsuperscript{3+} (KGW:Yb\textsuperscript{3+}) crystal.\cite{Bowman2002CLEO} Shortly thereafter, detailed pump-wavelength stability/gain fluctuation modeling\cite{Bowman2002JQE}, coupled to the development of a thermodynamic RBL formalism\cite{Mungan2003JOSAB}, motivated demonstrations of athermal 490-W  KGW:Yb\textsuperscript{3+} disk\cite{Bowman2005JQE} and 500-W YAG:Yb\textsuperscript{3+} rod lasers\cite{Bowman2010JQE}, both operating in quasi-CW modes. The success of high-power laser designs using thin-disk gain media\cite{Giesen1994} motivated theoretical\cite{Nemova2014OC} and experimental\cite{Yang2019OE} implementations of RBLs in Yb\textsuperscript{3+}-doped thin-disk geometries. This led to the realization that input pump and output laser modes needed to be matched to avoid thermal gradients in the gain medium.\cite{Mungan2003JOSAB,Nemova2014OC,Yang2019OE} These efforts culminated in a full analysis of optimal RBL performance.\cite{Sheik-Bahae2020JQE} Parallel efforts on gain-material platforms that naturally avoid thermal parasitics include theoretical analyses\cite{Nemova2009JOSAB,Mafi2020JOSAB} as well as experimental demonstrations of RBAs\cite{Knall2021PRL} and RBLs\cite{knall2021radiation} in single-mode optical fibers (Yb\textsuperscript{3+}-doped fused silica) and microspheres.\cite{Xia2021SPIE}

Other RBL advances include the development and optimization of Yb\textsuperscript{3+}-doped YLF, LuLiF$_4$ (LLF)\cite{Volpi2021SPIE}, YAG, and KY(WO$_4$)$_2$ (KYW:Yb\textsuperscript{3+})\cite{Cheng2023PhD} hosts, as well as proposals for tandem material systems, including semiconductor/YLF:Yb\textsuperscript{3+}\cite{Khurgin2020JOSAB} and mid-infrared YLF:Tm\textsuperscript{3+}/YLF:Ho\textsuperscript{3+} RBLs.\cite{rostami2019tm} There have also been promising cooling results reported for YLF:Tm\textsuperscript{3+} and BaY$_2$F$_8$:Tm\textsuperscript{3+} (BYF:Tm\textsuperscript{3+}) systems\cite{Rostami2020UNM} along with analogous advances with fiber-based RBLs, in terms of improved materials science\cite{Thomas2023OM} and design optimization.\cite{Chen2025SPIE} In whole, the above efforts point to the eventual realization of high-power, RBLs/RBAs in the near- and mid-infrared frequency regions, which will advance laser science as well as critical, low laser noise applications.  

\subsection{Metrology}\label{sect:metrology}
\subsubsection{Surface enhanced Raman spectroscopy}\label{subsect:SERS}
High precision sensing is another branch of applications that stands to benefit from laser refrigeration. As illustration, surface-enhanced Raman spectroscopy (SERS) provides molecular-level sensitivity and nanoscale temperature mapping of materials. In these measurements, photosensitive substrates or analytes are susceptible to photothermal damage. This stems from strong field confinement and absorption associated with field enhancing plasmonic nanostructures. Induced local heating can consequently photobleach dyes, denature proteins, or cause chemistry in fragile molecules.\cite{kneipp1997single} These thermal effects distort Raman spectral signatures, restrict excitation powers/acquisition times, and ultimately limit the Raman technique's quantitative reliability.

Numerous approaches have been explored to alleviate heating in SERS measurements.\cite{altangerel2023} This includes using pulsed illumination, thermal conduction layers, or cryogenic environments. These methods, however, are largely passive and do not dynamically counteract nanoscale temperature rises.

Integrating optical refrigeration into plasmonic systems offers an active and potentially transformative solution. This includes using \textnormal{Yb}$^{3+}$-, \textnormal{Er}$^{3+}$-, or \textnormal{Tm}$^{3+}$-doped YLF or NaYF$_4$ to cool plasmonic structures.\cite{roder2015laser,seletskiy2010laser,zhou2016laser} In such systems, heating and cooling coexist synergystically wherein plasmonic amplification of Raman scattering cross-sections is accompanied by RE\textsuperscript{3+}-induced cooling that reduces photothermal stress during measurements.

Early studies of laser cooling-assisted SERS platforms now report measurable temperature reductions and improved spectral stabilities under continuous-wave illumination.\cite{pant2020solid} Beyond mitigating sample degradation, such active cooling stands to refine the accuracy of Raman thermometry by providing nanoscale temperature control of specimens. With continued advances in RE\textsuperscript{3+}-doped materials and plasmonic design, future self-cooling SERS systems may enable long-duration, high-power Raman spectroscopy of sensitive biological and molecular systems without external cryogenics.

\subsubsection{Optomechanics}\label{subsect:optomechanics}
Another area that will benefit from solid-state laser cooling is levitated optomechanics. In these studies, heating induced by high-power trapping lasers, leads to instability and increased noise levels, especially at low pressures. An encouraging development has therefore been the $\Delta T=59$ K laser refrigeration of individual $\alpha$-NaYF:Yb\textsuperscript{3+} nanoparticles, levitated by a single beam trap in vacuum. This holds promise for studying optically-levitated nanomaterials at lower pressure with intrinsic temperature control to minimize noise levels.\cite{luntz2021laser} Spherical CaF$_2$:Yb$^{3+}$ microparticles trapped in dual beam vacuum levitation setups similarly represent promising candidate systems.\cite{gregoire2025exploring}

\subsubsection{Actively cooled sensors}\label{subsubsect:actively_cooled_sensors}
Composite laser refrigeration/sensor materials have additionally been developed and demonstrated. Specific examples include nanodiamonds with nitrogen vacancy (NV) centers deposited atop YLF:Yb$^{3+}$ crystals. By optically cooling YLF:Yb$^{3+}$ $\Delta T=27$ K below room temperature, thermally-induced spectral wandering of nanodiamond NV centers has been suppressed.\cite{pant2020solidb} Similarly, YLF:Yb$^{3+}$ microcystals have been used to cool CdS optomechanical resonators by over 20 K.\cite{pant2020solid} This opens the door to potential applications in scanning probe microscopy, force sensing, and the ultraprecise measurement of atomic masses.

Finally, nanodiamonds shelled with $\alpha$-NaYF:Yb$^{3+}$ have been developed for applications in optically-levitated, high precision quantum sensing.\cite{forberger2022synthesis} To date, these composites have not refrigerated. These early-stage efforts in laser-refrigerated, levitated optomechanics and optically-cooled sensor composite materials, however, show promise and point to many fields of metrology being enhanced by micro- or nanoscale laser cooling materials.

\subsubsection{Gravitational waves and extremely narrow-linewidth lasers}\label{subsubsect:frequency_combs}
The vibration-free nature of optical refrigeration likewise makes it attractive for applications in high-precision sensing, which require cooling without mechanical vibrations that accompany the operation of mechanical pumps or the evaporation of liquid cryogens. As an illustration, a recent proposal for detecting high-frequency gravitational waves relies on the optical levitation of two-dimensional, disc-shaped particles within the arms of a Michelson interferometer.\cite{aggarwal2022} The two-dimensional morphology of the discs enhances forward-scattering of laser radiation, which increases the interferometer's sensitivity. Laser heating of the disc, caused by absorption of infrared levitation laser light, however, increases photon shot-noise from blackbody radiation. This decreases the interferometer's sensitivity. Recently, hexagonal $\beta$-NaYF disks with micron-scale diameters have been optically levitated in vacuum within two standing-wave trap arms of a Michelson interferometer.\cite{winstone2022} These discs have also been demonstrated to undergo laser cooling\cite{Felsted2022} and point to the eventual mitigation of blackbody photon shot noise in these measurements via laser cooling.

Of similar note is the recent realization of an ultrastable laser\cite{Kessler2012NaPho}, achieved by locking it to an external, cryogenically-refrigerated, monolithic Si cavity. The etalon is cooled to 124 K, where Si possesses a null coefficient of thermal expansion. This dramatically reduces cavity thermomechanical noise and imparts unprecedented frequency stability to the locked laser. Although, cooling was achieved using a vibration-stabilized, nitrogen gas cryostat, future studies will likely use vibration-free optical cryocoolers once cooling powers of approximately 0.2 W have been realized (see \textbf{Section \ref{subsubsect:power_scaling}}).

\section{Reproducibility and data deposition}\label{section:reproducibility}
\subsection{Rationale for reporting standards}
Because solid state laser cooling is premised on the delicate and precarious balance between radiative and non-radiative processes in materials (\textbf{Section \ref{subsect:cooling_formulas}}), it is exquisitely sensitive to small variations in material quality, optical coupling, heat management and thermometry.\cite{hasselbeck2007effect} This has led to questions about the reliability of condensed phase laser cooling claims over the years\cite{finkeissen1999cooling,Clark1996laser,mungan1996comment}, most recently within the area of semiconductor cooling.\cite{Zhang2013Laser,morozov2019can,nemova2010laser}

\begin{figure}
\centering
\includegraphics[width=0.65\linewidth]{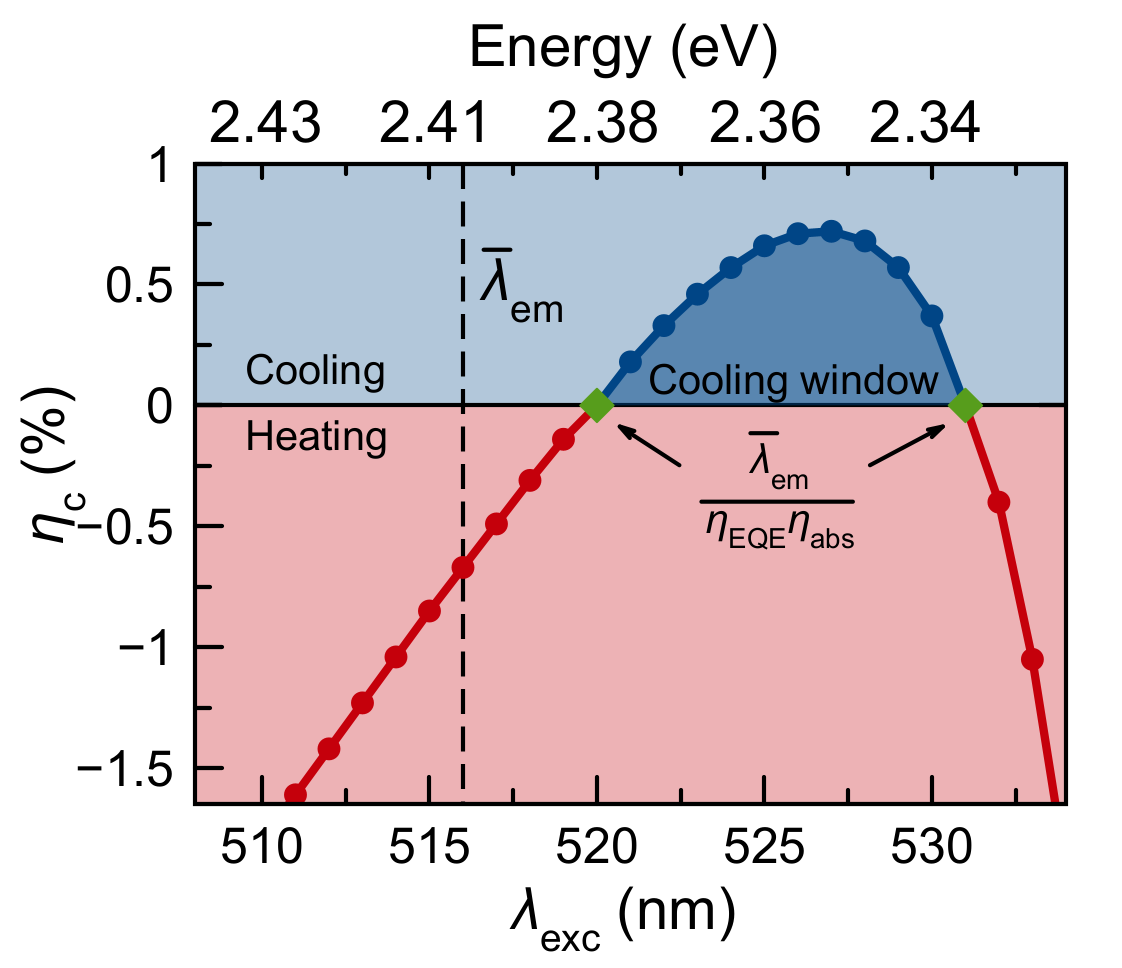}
\caption{Simulated (a) $\eta_\textrm{c}$ and (b) $\eta_\textrm{abs}$ for CsPbBr$_3$ nanocrystals ($\eta_\textrm{EQE}=0.991 $, $\alpha_\textrm{b}=3\times10^{-4}$ cm$^\textrm{-1}$). Indicated are cooling and heating regimes as well as corresponding transition points.}
\label{fig:Figure_4}
\end{figure}

The ongoing debate regarding semiconductor optical refrigeration highlights the need for a standardized, experimental minimum reporting criterion to ensure the reliability and reproducibility of cooling claims. Reference \cite{Zhang2025Principles} summarizes this criterion whose key points are outlined below.
\begin{itemize}
\item{\textbf{$\eta_{\textrm{EQE}}$ and $\eta_{\textrm{abs}}$:}} Most important is that requisite unity/near-unity $\eta_\textrm{EQE}$- and $\eta_\textrm{abs}$-values be measured and reported. For the former, absolute quantum yield measurements are preferred given complications with less direct (relative) approaches. These complications have been detailed in Reference \cite{zhang2019progress}. For the latter, accurate estimates of RE\textsuperscript{3+} or semiconductor $\alpha_\textrm{b}$-values are required. Alternatively, more involved temperature- and $\Delta E$-dependent up-conversion measurements can be used to estimate $\eta_\textrm{abs}$.\cite{zhang2024resonant}

\item{\textbf{Timescales:}} Observed $\Delta T$, $T_\textrm{final}$, and associated cooling timescales, $\tau$, must be reported. This is because thermodynamic constraints due to a specimen's effective heat capacity, including all potential heat loads from a substrate or surrounding matrix as well as environmental blackbody radiation, dictate realizable $\Delta T$, $T_\textrm{final}$, and involved timescales.

Temperature changes follow $T(t)\approx T_0-\frac{P_\textbf{cool}}{\kappa}(1-e^{-\frac{t}{\tau}})$ where $T_0$ is the initial temperature (K) and $\kappa$ is a heat transfer coefficient (W K$^{-1}$). An associated final temperature is therefore $T_{\textrm{final}} = \frac{P_\textbf{cool}}{\kappa}$ with an associated cooling (or heating) timescale of $\tau=\frac{C}{\kappa}$. $C$ is the specimen's effective heat capacity (J K$^{-1}$) and includes contributions from all thermal loads on the cooling material.

\item{\textbf{Thermometry:}} Details of the temperature measurement approach, as well as calibration procedure, must be provided. In general, specimen-independent measurement techniques such as thermal cameras are preferred (\textbf{Section \ref{subsect:thermometry}}). Primary thermometry should be cross-checked using a second, independent approach. For example, calibrated DLT thermometry could be used to confirm measured temperatures. However, it should be kept in mind that PL-based techniques such as DLT\cite{imangholi2006differential,patterson2010measurement,seletskiy2010laser} are susceptible to influence by extrinsic or intrinsic sources of spectral changes\cite{hasselbeck2007effect}, as described in \textbf{Section \ref{subsect:thermometry}}.

\item{\textbf{Cooling and Heating:}} Finally, reversible cooling and heating must be demonstrated. This is motivated by the need to exclude extraneous changes to materials, which mimic cooling. Here, \textbf{Equation \ref{eqn:eta_cool_equation_wavelength}} and \textbf{Figure \ref{fig:Figure_4}} reveal two heating/cooling transition points as $\lambda_\textrm{exc}$ is increased above $\bar\lambda_\textrm{em}$. Both transition points occur when $\eta_\textrm{c}=0$ (\textbf{Equation \ref{eqn:eta_cool_equation_wavelength}}) where $\lambda_\textrm{exc}=\frac{\bar\lambda_\textrm{em}}{\eta_\textrm{EQE}\eta_\textrm{abs}}$. Between these points, there exists a cooling window where $P_\textrm{cool}>0$ ($\eta_\textrm{cool}>0$). Outside the window, $P_\textrm{cool}<0$ ($\eta_\textrm{cool}<0$) so that only heating is observed. Studies should therefore demonstrate reversible cooling and heating about these points via control of $\lambda_\textrm{exc}$. One approach that readily satisfies this requirement and which simultaneously provides independent estimates of $\eta_\textrm{EQE}$ is LITMoS.\cite{melgaard2014identification,seletskiy2013precise}
\end{itemize}

\subsection{Reporting and data deposition checklists}\label{subsect:reporting_standards}
Reporting entails a comprehensive description of the methods used to acquire cooling data. This includes the following recommended reporting checklist.
\subsubsection{Specimen-related checklist}
\begin{itemize}
\setlength{\itemsep}{0.6pt}
\item  For RE\textsuperscript{3+}-doped materials, report specimen doping levels/concentrations and material purity.
\item For either RE\textsuperscript{3+}-doped or semiconductor nanostructures, provide information on crystallinity, mean size, ensemble polydispersity, and nature of the surface passivation. This includes nanostructure concentrations in systems that involve ensembles or dispersions.
\item Provide cooling specimen details (\textit{e.g.}, substrate, specimen supports, thermal isolation) as well as origin of parasitic heat loads.
\end{itemize}
\subsubsection{Laser cooling-related checklist}
\begin{itemize}
\setlength{\itemsep}{0.6pt}
\item Report $\eta_\textrm{EQE}$, details on how it was measured (\textit{e.g.}, integrating sphere or otherwise), and measurement errors.
\item Report $\eta_\textrm{abs}$ and/or $\alpha_\textrm{b}$. Ensure that the product, $\eta_\textrm{EQE}\eta_\textrm{abs}$, exceeds a critical cooling metric, $\eta_\textrm{EQE}\eta_\textrm{abs}>1-{\frac{\Delta E}{\bar E_{\textrm{em}}}}$.
\item Provide demonstrations of explicit cooling versus heating, including reversible transitions between the two regimes about $\frac{\bar\lambda_\textrm{em}}{\eta_\textrm{EQE}\eta_\textrm{abs}}$ See \textbf{Equation \ref{eqn:eta_cool_equation_wavelength}} and \textbf{Figure \ref{fig:Figure_4}}.
\item Provide measurement details of all temperature thermometries employed, including information on their calibration as well as temperature and time resolution.
\item Report estimated cooling/heating $\tau$ based on thermodynamic constraints on the system. Provide estimated $\Delta T$ and $T_\textrm{final}$. Values should be accompanied by explicit comparisons to experiment.
\item Detail employed excitation sources, including information on spectral linewidths, wavelength tuning method, output polarization, output power, power stability, and beam characteristics at the sample, \textit{e.g.}, ${1\over e^2}$ diameter.
\end{itemize}

\section{Solid state cooling limitations}\label{section:limitations} 
\subsection{Minimum achievable temperature}\label{subsubsect:MAT_and_gMAT}
For YLF:Yb\textsuperscript{3+}, the workhorse material of bulk optical refrigerators, its gMAT is $\approx$70 K for a 10\% doping concentration.\cite{volpi2019optical} This cooling floor primarily stems from its non-zero, background absorption coefficient of $\alpha_\mathrm{b} = 2\times 10^{-4}\ \rm{cm}^{-1}$ at room temperature, which decreases with decreasing temperature.\cite{volpi2019optical} Transition-metal ions such as iron are considered to be the primary contributor to $\alpha_\textrm{b}$.\cite{hehlen2007model,melgaard2014identification} Although increasing Yb\textsuperscript{3+} concentrations can compensate for $\alpha_\textrm{b}$, realizing net cooling with Yb\textsuperscript{3+} doping values larger than 20\% remains challenging due to pronounced energy transfer to said impurities.\cite{puschel2024impact}

Reducing YLF:Yb\textsuperscript{3+} impurity levels is therefore key to achieving a lower YLF:Yb\textsuperscript{3+} gMAT-value. To this end, Patterson \textit{et al.} have reported preparing ultra pure metal fluorides by combining chelate-assisted solvent extraction with HF-gas fluorination. What results is a 1000-fold reduction of transition-metal contamination to the 100 part-per-billion (ppb) level.\cite{patterson2011case} Repetitive crystallization processes, such as zone-refining or zone-melting also represent promising techniques for eliminating impurities with low segregation coefficients.

Despite above noted YLF:Yb\textsuperscript{3+} milestones (\textbf{Table \ref{table:RE3+_crystal_milestones}}), its predominance stems from the availability of high-quality, high-purity single crystals rather than any intrinsic advantage it possesses over other Yb\textsuperscript{3+}-doped crystals. We have already indicated KY\textsubscript{3}F\textsubscript{10}:Yb\textsuperscript{3+} as a promising alternative material, which possesses a lower gMAT.\cite{puschel2022laser, puschel2023ytterbium} Using different RE\textsuperscript{3+} ions with lower $\bar E_\textrm{em}$-values than Yb\textsuperscript{3+} (\textbf{Table \ref{table:RE3+_summary}}) may further reduce RE\textsuperscript{3+} gMAT values. Here, Rostami and Sheik-Bahae have identified BaY\textsubscript{2}F\textsubscript{8}:Ho\textsuperscript{3+} as a candidate system for 2-$\mu$m optical refrigeration with the potential for a sub 70 K gMAT.\cite{rostami2021observation}

For cooling nanocrystals, the Purcell effect, which enhances spontaneous emission rates in resonant optical cavities or photonic structures, can substantially increase $\eta_\textrm{c}$ and, by extension, reduce associated gMAT-values.\cite{ju2024purcell,benzaouia2024purcell} A numerical analysis by Peng \textit{et al.} reveals a YLF:Yb\textsuperscript{3+} nanocrystal gMAT of 38 K for a Purcell factor of 180 along with realistic $\eta_\mathrm{EQE}$ and $\eta_\mathrm{abs}$-values.\cite{ju2024purcell} This assumes no additional losses are introduced, as parameterized by $\alpha_{\textrm{b}}$.

\subsection{Power scaling}\label{subsubsect:power_scaling}
An important challenge for any optical cryocooler is to increase its cooling power. Current best of class RE\textsuperscript{3+} cryocoolers have $P_\textrm{cool}$-values of order 0.1 W.\cite{seletskiy2010laser} Although \textbf{Equation \ref{eqn:Pcool_equation_wavelength}} points to $P_\textrm{cool}$ increasing linearly with increasing $P_\textrm{abs}$, this only applies at intensities much smaller than the saturation intensity of RE\textsuperscript{3+} resonant transitions. Above saturation, $\eta_\textrm{abs}$ and $\eta_\textrm{c}$ decrease and, in turn, lower $P_\textrm{cool}$ ($\alpha_\textrm{b}$ is assumed to not saturate). 

Current astigmatic Herriott cell\cite{Gragossian2016OE} multi-pass pumping schemes minimize saturation effects by spreading laser passes throughout the whole specimen volume. This comes at the cost of increasing the complexity of Herriott cell engineering. Another strategy reduces saturation effects by using total internal reflection to trap pump light at vacuum/crystal interfaces.\cite{thiede2005cooling} This is done by coating cooling elements with highly reflective mirrors. A drawback of the approach, though, is a reduced absorption efficiency, due to sub-optimal orientation of internally reflected pump light polarizations relative to a cooling element's optical axis.

Increasing specimen size also effectively lowers pump intensities. Unfortunately, the approach increases the likelihood of detrimental PL reabsorption, which redshifts $\bar\lambda_\textrm{em}$ to reduce $\eta_\textrm{c}$ (\textbf{Equation \ref{eqn:eta_cool_equation_wavelength}}). For RE\textsuperscript{3+} cryocoolers, large or oddly-shaped crystals further impede good thermal conductivity to payloads (\textbf{Section \ref{subsect:cryocooler}}). Thus, in combination with above-outlined $P_\textrm{cool}$ enhancement strategies, the only current solution for scaling solid state optical refrigeration cooling powers involves parallelizing cooling by attaching multiple cooling elements to individual payloads. This increases the complexity of solid state laser cooling solutions and stands as an open challenge for the field going forward.

\section{Outlook}\label{section:outlook}
Solid-state laser refrigeration remains a vibrant and dynamic research field with wide-ranging opportunities for breakthroughs in the design and discovery of materials with unity/near-unity $\eta_\textrm{EQE}$- as well as with correspondingly large $\eta_\textrm{abs}$-values. Attesting to this, advances in the synthesis of high-quality, RE\textsuperscript{3+}-doped materials have enabled important demonstrations of payload cooling to cryogenic temperatures.\cite{hehlen2018,kock2022ol}

Analogous advances in the synthesis of RE\textsuperscript{3+}-doped ceramic nano- and micron-scale materials promise important breakthroughs in the solid-state laser refrigeration of dielectrics used in quantum communication and quantum sensing\cite{gali2019}, the generation of quantum squeezed states of light\cite{magrini2022}, high-power optically-levitated accelerometers\cite{gregoire2025exploring}, the detection of high-frequency gravitational waves\cite{aggarwal2022,winstone2022}, surface-enhanced Raman spectroscopy\cite{altangerel2023}, and also the design and realization of radiation balanced microlasers and amplifiers.\cite{knall2021radiation, fb2018} Additionally, based on pioneering demonstrations of cold Brownian motion at physiological conditions\cite{roder2015laser}, the in vitro laser cooling of organelles, ion channels, and molecular machines represents a promising future direction for single-molecule biophysics.\cite{peng2025}

For all above use cases, semiconductor optical refrigeration represents a significant advance given lower predicted cooling floors. However, despite decades of efforts, their laser cooling remains a significant challenge. An ongoing debate exists as to whether actual semiconductor cooling examples exist.\cite{pant2019,morozov2019can} Here, we identify low-dimensional semiconductors as a promising future direction of research towards realizing verifiable examples of semiconductor optical refrigeration.

Key challenges remain in improving the overall thermodynamic efficiency of solid-state laser cooling. ASPL photon recycling is one viable approach for boosting thermodynamic efficiencies closer to the Carnot limit.\cite{epstein2009optical} Proposed designs for scaling cooling power already approach the performance requirements of various applications discussed in this Primer. In parallel, the development of narrow-gap materials in the telecom ($\sim 1.5$ $\mu$m) or mid-infrared (2.5-5 $\mu$m) spectral ranges represents another promising pathway for increasing thermodynamic efficiencies. With these enhancements, solid-state laser coolers are poised to enable multiple uses, from basic and applied research to industrial applications. 

\printbibliography 

\section{Acknowledgments}\label{section:Acknowledgments}
Y.D., S.Z., A.R.A., M.P.H., P.J.P., D.V.S., and M.K. thank the Air Force Office of Scientific Research for financial support via the MURI:MARBLe project (FA9550-16-1-0362) and Grant FA9550-24-1-0668. We dedicate this work to Professor Mansoor Sheik-Bahae (1956-2023), a pioneer in the field of solid-state laser cooling who laid the foundations of many theoretical insights, modeling results, and experimental ideas, summarized in this Primer.

\section{Author information}\label{section:AuthorInfo}
\subsection*{Authors and Affiliations}
\paragraph{\textbf{Department of Chemistry and Biochemistry, University of Notre Dame, Notre Dame, IN, USA.}}
\noindent Yang Ding, Shenghao Zhang \& Masaru Kuno

\paragraph{\textbf{Department of Physics and Astronomy, University of New Mexico, Albuquerque, NM, USA.}}
\noindent Alexander R. Albrecht \& Denis V. Seletskiy

\paragraph{\textbf{Department of Materials Science, University of Washington, Seattle, Washington, USA.}}
\noindent Zhaojie Feng, Lars Forberger \& Peter J. Pauzauskie

\paragraph{\textbf{Leibniz-Institut für Kristallzüchtung (IKZ), Berlin, Germany.}}
\noindent Hiroki Tanaka

\paragraph{\textbf{Los Alamos National Laboratory, Los Alamos, NM, USA.}}
\noindent Markus P. Hehlen

\paragraph{\textbf{Department of Engineering Physics, Polytechnique Montréal, Montreal, Canada.}}
\noindent Galina Nemova \& Denis V. Seletskiy

\paragraph{\textbf{Department of Physics and Astronomy, University of Notre Dame, Notre Dame, IN 46556, USA}}
\noindent Masaru Kuno

\subsection*{Contributions}
All authors contributed to each section, and reviewed and edited the manuscript before submission.

\section{Ethics declarations}
\subsection*{Competing interests}
The authors declare no competing interests.

\end{document}